\documentstyle[prl,aps,preprint]{revtex}

\let\ssection=\section
\renewcommand{\section}{\setcounter{equation}{0}\ssection}

\begin{document}
\draft
\title{\bf On physical foundations and observational 
effects of cosmic rotation}
\bigskip
\author{\sl Yuri N. Obukhov}
\maketitle
\centerline{\it Department of Theoretical Physics,}
\centerline{\it Moscow State University, 117234 Moscow, Russia}

\begin{abstract}
An overview of the cosmological models with expansion, shear and
rotation is presented. Problems of the rotating models are discussed, their
general kinematic properties and dynamical realizations are described. A
particular attention is paid to the possible observational effects which
could give a definite answer to the question: is there evidence for the 
rotation of the universe? 
\end{abstract}


\section{Introduction}

In his lecture {``Spin in the Universe"} \cite{whit}, E.T. Whittaker 
has drawn attention to the following questions: {\it ``Rotation is a universal 
phenomenon; the earth and all the other members of the solar system rotate 
on their axes, the satellites revolve round the planets, the planets revolve 
round the Sun, and the Sun himself is a member of the galaxy or Milky Way 
system which revolves in a very remarkable way. How did all these rotary 
motions come into being? What secures their permanence or brings about their 
modifications? And what part do they play in the system of the world?''}

 The standard Friedman-Lema\^{\i}tre cosmological models are based on very
 idealized assumptions about the spacetime structure and matter source. 
 The Friedman-Robertson-Walker metrics describe an exactly isotropic and
 homogeneous world filled with an ideal fluid (radiation and dust are the
 well known particular cases). Although these assumptions are qualitatively 
 supported by observations, it seems doubtful that the geometry of the
 universe and its physical contents are so finely tuned starting with the
 very Big Bang till the modern stage of the evolution. A reasonable step
 is to look for more general and realistic models with the assumptions of
 isotropy, homogeneity, and ideal fluid matter contents, relaxed. Here we 
will not discuss inhomogeneous cosmologies which are reviewed, e.g., in
\cite{mac1,mac2}. 

Already in the early years of the general relativity theory, attempts
were made to construct (exact or approximate) solutions of the gravitational
field equations with rotating matter sources (see, e.g. 
\cite{lanc,a2,a3,lewis,van}). Lanczos \cite{lanc} appears to be the
first who considered the universe as the largest possible rotating physical 
system in a model of a rigidly rotating dust cylinder of an infinite radius. 
Dust density in this solution (which was later rederived by van Stockum 
\cite{van}) diverges at radial infinity, which was a serious problem.

In 1946, Gamow \cite{gamov} stressed a possible significance of the cosmic 
rotation for explaining the rotation of galaxies in the theories of galaxy 
formation, see also his discussion in \cite{gamov2}. Independently, 
Weizs\"acker \cite{weiz1} initiated the study of the primordial turbulence 
and of its role in the formation of the universe's structure. Soon after this, 
K. G\"odel \cite{goe1} had proposed a stationary cosmological model with 
rotation in the form (coordinates $\{t,x,y,z\}$, $a=$const):
\begin{equation}
ds^2 = a^2\left(dt^2 - 2e^{x}dtdy + {1\over 2}e^{2x}dy^2 - dx^2 - dz^2
\right).\label{gd}
\end{equation}
In the G\"odel's model, the matter is described as dust with the energy 
density $\varepsilon$, and the cosmological constant $\Lambda$ is nontrivial 
and negative (i.e. its sign is opposite to that introduced by Einstein). The 
angular velocity $\omega$ of the {\it cosmic rotation} in (\ref{gd}) is given 
by $\omega^2 = {1\over 2a^2}= 4\pi G\varepsilon = - \Lambda$, with $G$ as 
Newton's gravitational constant. For many years this model became a 
theoretical ``laboratory" for the study of rotating cosmologies, see e.g.
\cite{a1,buch,bampi,coh,his,hoens,kundt,pfarr,pim1,pim2,stein,wright2}. 
The stationary solution (\ref{gd}) is distinguished among other spatially
homogeneous models: the so-called G\"odel theorem (complete proof is given
in \cite{ozv1}) states that the only spatially homogeneous models with 
zero expansion and shear are Einstein's static universe and G\"odel model
(\ref{gd}). 

Subsequently, G\"{o}del \cite{goe2} himself outlined a more physical {\it 
expanding} generalizations of (\ref{gd}), although without giving explicit 
solutions. Later a considerable number of exact and approximate models with 
rotation and with or without expansion were developed. At the end of this
paper one can find a list of references which may be considered as a
replacement of a detailed review of the previous work on rotating cosmology.

Here we would like to mention only briefly some earlier papers. Maitra 
\cite{maitra} gave a generalization of van Stockum's solution which is 
cylindrically symmetric inhomogeneous stationary dust-filled world with 
rotation and shear. Maitra formulated a simple criterion of absence of the
closed timelike curves. Wright \cite{wright1} presented another inhomogeneous
cylindrically symmetric solution for dust {\it plus} cosmological term. 

Dust and cosmological constant represented the matter source in the
series of papers by Sch\"ucking and Ozsv\'ath 
\cite{osch1,osch2,osch3,osch4,osch5,ozv1,ozv2,ozv3,ozv4,ozv5}, where the
spatially homogeneous models with expansion, rotation and shear were 
considered. The closed Bianchi type IX solution (\cite{osch5} with 
stationary limit \cite{osch1,osch2}) has drawn a special attention. 
Most general anisotropic Bianchi type IX model was studied by Matzner 
\cite{mat1,mat2} who took ideal fluid plus collisionless massless particles 
(photons and neutrinos) as matter source. The approximate solution was given.
Later, Rebou\c{c}as and de Lima \cite{reb4} described the exact analytic 
Bianchi type IX solution with expansion, shear and rotation for a source 
represented by dust with heat flow and cosmological constant. Analogously, 
Sviestins \cite{sve2} presented an exact Bianchi type IX solution for ideal
fluid plus heat flow. In an interesting recent paper, Gr{\o}n \cite{gron2} 
gave a {\it shear-free} Bianchi type IX solution in absence of matter (but
with cosmological constant) which can be considered as a rotating 
generalization of de Sitter spacetime. In fact, the latter is recovered
asymptotically after the quick decay of the vorticity.

The wide spectrum of spatially homogeneous models includes approximate and
exact solutions for all Bianchi types and different matter contents (from
simple dust to sometimes quite exotic energy-momentum currents). To mention
but a few, Ruzmaikina-Ruzmaikin \cite{ruz}, Demia\'nski-Grishchuk \cite{dem},
and Batakis-Cohen \cite{bac} presented approximate solutions of Bianchi 
type V and VII with expansion, shear and vorticity for the ideal fluid in
non-comoving coordinates. Bradley and Sviestins \cite{bs} added heat flow
to the ideal fluid source in their analysis of a Bianchi type VIII model.
In a series of papers, Fennelly \cite{fen2,fen3,fen4,fen5,fen6}
has studied Bianchi models of different types for homogeneous universes
filled with perfect charged (conducting) fluid and magnetic field. A neutral
ideal fluid plus electromagnetic field and null radiation represented the
matter source in a shear-free expanding and rotating solution of Patel-Vaidya 
\cite{pat3} which can be interpreted as a standard Friedman-Robertson-Walker 
cosmology with rotating matter \cite{gron5}. More sophisticated matter source
may include viscous fluid and scalar fields, see, e.g., recent papers
\cite{mani1,mani2,mani3,mani4,mani5,mani6,mani7,mani8,mani9,mani10,mani11}.
More exact solutions, which we will not discuss in detail, 
\cite{aga1,aga2,aga3,bac,bs,dray,n1,rt,reb1,reb2,ros1,sve2,soares} 
were reported in the literature. Most of these models were overviewed by
Krasi\'nski \cite{kras1,kras2,kras3,kras4,kras5}, more recent review of 
generalized cosmologies with shear and rotation see in 
\cite{kras8,kras9,kras10}. 

It is worthwhile to notice that along with the constant and deep interest 
in the rotating cosmologies, historical development has revealed several 
difficulties which were considered by the majority of relativists as the 
arguments against the models with nontrivial cosmic rotation. One of the 
problems was the the stationarity of G\"odel's model (\ref{gd}). Apparent 
expansion of the universe is usually related to the fact of the red shift 
in the spectra of distant galaxies, and thus all the standard cosmological 
models are necessarily non-stationary. It was discovered, however, that it 
is impossible to combine pure rotation and expansion in a solution of the 
general relativity field equations for a simple physical matter source, such 
as a perfect fluid \cite{ellis1,ellis2,ellis3}. There are two ways to overcome 
that difficulty: one should either take a more general energy-momentum or to
add cosmic shear, the explicit examples were already mentioned above. The 
possibility of combining cosmic rotation with expansion was the first 
successful step towards a realistic cosmology. 

Another problem which was discovered already by G\"odel himself, was 
related to causality: the spacetime (\ref{gd}) admits closed timelike
curves. This was immediately recognized as an unphysical property because 
the causality may be violated in such a spacetime, see relevant discussion
in \cite{goe3,hawell,guts1,guts2,lath,pfarr,stein}. Considerable efforts
were thus focused on deriving completely causal rotating cosmologies. In his
last work devoted to rotation, G\"odel without proof mentions the possibility
of positive solution of the causality problem \cite{goe2}. First explicit
solutions were reported later \cite{maitra,osch3}. Maitra \cite{maitra}
formulated a simple criterion for the (non)existence of closed timelike 
curves in rotating metrics which we will discuss below. Eventually, it
became clear that the breakdown of causality is not inevitable consequence
of the cosmic vorticity, and many of the physically interesting rotating 
models were demonstrated to be completely causal. That was the
second important step in the development of the subject.

The discovery of the microwave background radiation (MBR) has revealed the 
remarkable fact that its temperature distribution is isotropic to a very 
high degree. This fact was for a long time considered as a serious argument 
for isotropic cosmological models and was used for obtaining estimates on 
the possible anisotropies which could take place on the early stages of the 
universe's evolution. In particular, homogeneous anisotropic rotating 
cosmologies were analyzed in \cite{haw,colhaw,bjs,temp1,kogut,wolfe}, and 
very strong upper limits on the value of the cosmic rotation were reported. 
Unfortunately, usually in these studies the effects of vorticity were not 
separated properly form the cosmic shear.\footnote{Even now in the literature 
one can often encounter statements like: ``If the universe had 
vorticity, the microwave background would be anisotropic..." \cite{sol1}.} 
For the first time, in \cite{banach,2}, the effects of pure cosmic rotation 
were carefully separated from shear effects. The most significant result is 
the demonstration that in a large class of spatially homogeneous rotating 
models the pure rotation leads neither to an anisotropy of the MBR temperature 
nor to a parallax effects, although the geometry remains completely causal. 
A typical representative of that class is the so called G\"odel type metric 
with rotation and expansion (\ref{met1}) which is considered below. Most 
recent analyzes, based also on the COBE data on the anisotropy of MBR, see 
\cite{bunn,mes,kogut}, again estimated not a pure rotation but rather a shear. 

The last but not least problem is the lack of direct observational evidence 
for the cosmic rotation. Attempting at its experimental discovery one should
study possible systematic irregularities in angular (ideally, over the whole
celestial sphere) distributions of visible physical properties of sources 
located at cosmological distances. Unfortunately, although a lot of data is 
already potentially accumulated in various astrophysical catalogues, no
convincing analysis was made in a search of the global cosmic rotation. 
Partially this can be explained by the insufficient theoretical study of 
observational cosmology with rotation. To our knowledge, till the recent 
time there were only few theoretical predictions concerning the possible 
manifestations of the cosmic rotation. Mainly these were the above mentioned 
estimates of MBR anisotropies \cite{haw,colhaw,temp1,kogut} (also of the 
$X$-ray background anisotropies \cite{wolfe}) and the number counts  
analyses \cite{lanc,goe2,wesson,mavr,fen}. Besides that, it was pointed out 
in \cite{tre1,rub} that, in general, parallax effects can serve as a 
critical observational test for the cosmic rotation.

Very few purely empirical analyses (without constructing general relativistic 
models) of the angular distributions of astrophysical data are available which 
interpreted the observed systematic irregularities as the possible effects of 
rotation. Of the early studies, let us mention the reports of Mandzhos and
Tel'nyuk-Adamchuk \cite{mand1,mand2,mand3} on the preferential orientation of
clusters of galaxies, Valdes et al \cite{valdes} on the search of coherent
ellipticity of galaxy images, and Andreasyan \cite{andr1,andr2} on flattening
of distant galaxies.

Birch \cite{birch1,birch2,conw} (Jodrell Bank Observatory) reported on the 
apparent anisotropy of the distribution of the observed angle between 
polarization vector and position of the major axis of radio sources, and 
related this global effect (using heuristic arguments) to the nontrivial 
cosmic rotation. Subsequently, these observations were disputed from a 
statistical theory viewpoint \cite{phinney,kron1,kron2}, but the analysis 
using indirectional statistics \cite{kend} confirmed Birch's results. It seems
worthwhile to mention that the independently data from the Byurakan radio 
observatory \cite{andr1,andr2} indicated the same anisotropy along, 
approximately, the same direction. These observations have stimulated the
interest in rotating models; nevertheless, no further observations were 
reported until the most recently, when the new data appeared 
\cite{nod1,nod1,nod4,nod5} claiming the same effect. Note however, that 
again the statistical significance of observations is under discussion 
\cite{anod1,anod2,anod3,anod4,anod5}.

The theoretical analysis of the observational effects in the class of viable 
rotating cosmologies \cite{banach,jetp1,2,18} has shown that the cosmic 
rotation produces a typical dipole anisotropy effect on the polarization of 
electromagnetic waves. Considering the case of the distant radio sources, 
one finds the dipole distribution \cite{jetp1,18} $\eta=\omega_0\,r\,\cos
\theta$, see (\ref{final}) below, for the angle $\eta$ between the vector of
polarization and an observed direction of the major axis of the image of a 
radio source. Here $\omega_0$ is the present value of the cosmic rotation, 
$r$ is the (apparent area) distance to a source, and $\theta$ is the spherical 
angle between the rotation axis and the direction to the source, $Z$ is the 
red shift. The data of Birch and the new data of Nodland and Ralston 
\cite{nod1,nod1,nod4,nod5} may be extremely important for improving 
the direct estimates of vorticity. 

The study of observational effects in rotating and expanding cosmological
models is essentially based on the kinematical properties of spacetime. 
The next step is to investigate the {\it dynamical} aspects of cosmological 
models with rotation, in other words, to construct viable models as solutions 
of the gravitational field equations. The ultimate goal is to achieve a 
self-consistent description of the physical matter sources and to establish a 
complete scenario of a realistic evolution of the universe, including the most 
early stages of the big bang. Already Lema\^{\i}tre \cite{lem} and Whittaker 
\cite{whit} raised questions about the nature of a rotating ``primeval atom'' 
from which the universe is supposed to emerge. A related problem is the 
developments of matter irregularities which is relevant to the galaxy and 
large-scale structure formation. Gamov \cite{gamov,gamov2,gamov3} and 
Weizs\"acker \cite{weiz1} stressed the importance of the cosmic vorticity
in the early universe. Silk \cite{silk1,silk2} was the first to demonstrate 
instability of rotating models to axial density perturbations and stability 
to perturbations in the plane of rotation. The idea that primordial 
vortical motions of cosmological matter gave rise to the formation of
large-scale structure underlied the related study of the so-called whirl
theory (cosmological turbulence) developed some time ago by Ozernoy and
Chernin \cite{che1,che2,che3,che4,che5,kur1,kur2,kur3,kur4,kur5,ozer1}, 
see also \cite{nariai,oort,anile5,dalla1,dalla2,harri1,harri2,silk3}.
In the recent paper Li \cite{li} has demonstrated that the global cosmic 
rotation can provide a natural origin of the rotation of galaxies, and the
empirical relation $J\sim M^{5/3}$ between the angular momentum $J$ and the 
mass $M$ of galaxies can be derived from the cosmological vorticity. [See 
Brosche \cite{brosche1,brosche2,brosche3,brosche4} for empirical data, and 
Tassie \cite{tassie} for the string-theoretical discussion].

Of the recent theoretical developments, it is worthwhile to mention the 
work \cite{grish} in which Grishchuk has demonstrated that the rotational 
cosmological perturbation can be naturally generated in the early universe 
through the quantum-mechanical ``pumping'' mechanism similar to the primordial 
gravitational waves. These perturbations were considered as the source of 
the large-angular-scale anisotropy of MBR, and comparison with observations 
was made. However, as was recognized by Grishchuk, the pumping mechanism only 
works when the primeval cosmological medium can sustain torque oscillations. 
No attempt was made to study the possible physical nature of the relevant 
cosmological matter. Exploiting the close relation between rotation and spin, 
one can try to attack this problem using the models of continua with 
microstructure. Self-consistent variational general relativistic theory of 
spin fluids was constructed in \cite{flu1,flu2,flu3}. The investigation of the 
development of small perturbations for the models with spinning cosmological 
matter \cite{flu3} yielded a generalization and refinement of the result by
Silk \cite{silk2} about instability to perturbations along the spin.

Most of the dynamical realizations of rotating cosmological models were
obtained in the framework of general relativity theory as the exact or
approximate solutions of Einstein's gravitational field equations in four
dimensions. However, we can mention also the study of higher-derivative
extensions of general relativity \cite{acc1,acc2,acc3,acc4,hwang}, solutions
in 5 dimensions \cite{5dim,reb10}, in 3 dimensions \cite{gur,3dim1,3dim2}, 
and in string-motivated effective cosmology \cite{bar2,bar3}. 
On the other hand, rotation, spin and torsion are closely interrelated in the 
Poincar\'e gauge theory of gravity \cite{fwh3,fwh4,PBO,IPS,MieB}, and hence 
it is quite natural to study cosmologies with rotation within the gauge 
gravity framework. Exact rotating solutions in Einstein-Cartan theory of 
gravity with spin and torsion were described, e.g., in 
\cite{minn,bed1,tiom,fig,aman,4,flu1,pal1,pal2,panov11,smal1,smal2}.
General preliminary analysis of the separate stages of the universe's 
evolution was made in our works \cite{iko1,iko2,iko3}, while in \cite{4,io} 
the complete cosmological scenarios are considered.

Summarizing the present state of the problem, it is noteworthy once again to
quote Whittaker \cite{whit}: {\it ``It cannot be said, however, that any of 
the mathematical-physical theories that have been put forward to explain spin
(rotation) in the universe has yet won complete and universal acceptance;
but progress has been so rapid in recent years that it is reasonable to hope 
for a not long-delayed solution of this fundamental problem of cosmology''}.


\section{Stationary cosmological model of G\"odel type}\label{statged}

The G\"odel metric (\ref{gd}) represents a particular case of a wider
family of stationary cosmological models described by the interval
\begin{equation}
ds^2 = dt^2 - 2\sqrt{\sigma}e^{mx}dtdy -
(dx^2 + ke^{2mx}dy^2 + dz^2),\label{metS}
\end{equation}
where $m,\sigma,k$ are constant parameters. Clearly, $\sigma >0$, and
for definiteness, we choose $m > 0$. The metric (\ref{metS}) is usually 
called the model with rotation of the G\"odel type. Coordinate $z$ gives 
the direction of the global rotation, the magnitude of which is constant:
\begin{equation}
\omega = \sqrt{{1\over 2}\omega_{\mu\nu}\omega^{\mu\nu}}=
{m\over 2}\sqrt{\sigma\over {k+\sigma}}.\label{rotged0}
\end{equation}
As we see, vanishing of $m$ or/and $\sigma$ yields zero vorticity. 

Let us describe the geometry of a Riemannian spacetime (\ref{metS}) in
detail. The determinant of the metric is equal $(\det g) = - (k + \sigma)
e^{2mx}$, and hence we assume that $(k + \sigma) > 0$ in order to have
the Lorentzian signature. 

Technically, all calculations are simplified greatly when exterior
calculus is used. It is convenient to choose at any point of the spacetime 
(\ref{metS}) a local orthonormal (Lorentz) tetrad $h^{a}_{\mu}$ so that, as 
usual, $g_{\mu\nu}=h^{a}_{\mu}h^{b}_{\nu}\eta_{ab}$ with $\eta_{ab}=
{\rm diag}(+1,-1,-1,-1)$ the standard Minkowski metric. This choice is not 
unique, and we will use the gauge in which
\begin{equation}
h^{\hat{0}}_{0}=1,\quad h^{\hat{0}}_{2}=-\sqrt{\sigma}e^{mx},\quad
h^{\hat{1}}_{1}=h^{\hat{3}}_{3}=1,\quad h^{\hat{2}}_{2}=e^{mx}\sqrt{k+\sigma}.
\label{tetrad0}
\end{equation}
 Hereafter a caret denotes tetrad indices; Latin alphabet is used for the 
 local Lorentz frames, $a,b,...=0,1,2,3$. The local Lorentz coframe one-form
is defined, correspondingly, by
\begin{equation}
\vartheta^a := h^a_\mu\,dx^\mu.\label{cofr}
\end{equation}

The curvature two-form for (\ref{metS}) is computed straightforwardly: 
\begin{equation}
R^{ab} = \omega^2\,\vartheta^a\wedge\vartheta^b + {k\,m^2\over k + \sigma}
\left(\delta^a_{\hat 1}\delta^b_{\hat 2} -\delta^b_{\hat 1}\delta^a_{\hat 2}
\right)\vartheta^{\hat{1}}\wedge\vartheta^{\hat{2}} \qquad
{\rm for}\qquad a,b=0,1,2.\label{gedcurv}
\end{equation}
It is worthwhile to note that the spacetime (\ref{metS}) has a formal
structure of a product of the curved {\it three-dimensional} manifold with 
the curvature (\ref{gedcurv}) times flat one-dimensional space ($z$-axis),
which is reflected in the range of indices in (\ref{gedcurv}). 

The only nontrivial components (with respect to the local Lorentz
frame (\ref{tetrad0})) of the Weyl tensor are as follows:
\begin{equation}
C_{\hat{0}\hat{1}\hat{0}\hat{1}} = C_{\hat{0}\hat{2}\hat{0}\hat{2}} =
-\,C_{\hat{2}\hat{3}\hat{2}\hat{3}} = -\,C_{\hat{3}\hat{1}\hat{3}\hat{1}}
={1\over 2}\,C_{\hat{1}\hat{2}\hat{1}\hat{2}} = 
-\,{1\over 2}\,C_{\hat{0}\hat{3}\hat{0}\hat{3}} = {m^2\over 6}\left(
{k\over k + \sigma}\right).\label{weyl0}
\end{equation}
One thus verifies that the model (\ref{metS}) belongs to type $D$
according to Petrov's classification. 

The first dynamical realization of the model (\ref{metS}) was obtained
by G\"odel \cite{goe1}. In general relativity theory, metric satisfies 
Einstein's field equations:
\begin{equation}
G_{ab} := R_{ab} - {1\over 2}g_{ab}R = \,\Lambda\,g_{ab} + \kappa T_{ab}.
\end{equation}
G\"odel \cite{goe1} considered the simplest matter source: ideal dust with
the energy-momentum tensor $T_{ab}=\rho\,u_a\,u_b$. Substituting the
four-velocity of the co-moving matter $u^a=\delta^a_{\hat 0}$, and using
the components of the Einstein tensor,
\begin{equation}
G_{\hat{0}\hat{0}}=-\omega^2\left(1 + 4{k\over\sigma}\right),\quad
G_{\hat{1}\hat{1}}=G_{\hat{2}\hat{2}}=\omega^2,\quad
G_{\hat{3}\hat{3}}=\omega^2\left(3 + 4{k\over\sigma}\right),
\end{equation}
one finds the parameters of the famous G\"odel solution [compare with
(\ref{gd})]:
\begin{equation}
\omega^2 = -\,\Lambda, \qquad \kappa\rho=2\,\omega^2, \qquad 
{k\over\sigma} =-\,{1\over 2}. \label{gedpar}
\end{equation}
As we see, both the cosmological constant $\Lambda$ and the parameter
$k$ must be {\it negative}. 

Without confining ourselves to a particular dynamical realization,
we should study the whole class of stationary models (\ref{metS}). 
There is a wide symmetry group for that metric. Three evident isometries 
are generated by the Killing vector fields
\begin{equation}
 \xi_{(0)} ={\partial\over\partial t},\qquad  
 \xi_{(1)} ={\partial\over\partial y},\qquad  
 \xi_{(2)} ={\partial\over\partial z}.\label{iso0}
\end{equation}
These are all mutually commuting. However, one can find two more Killing
vectors. Most easily this can be done with the help of the suitable
coordinate transformations. 

Without touching $z$ coordinate, let us transform the rest three
$$
(t,\,x,\,y)\longrightarrow 
(\overline{\tau},\,\overline{r},\,\overline{\varphi})
$$
as follows:
\begin{eqnarray}
e^{mx}&=&e^{\overline{\varphi}}\cosh(m\overline{r}),\\
ye^{mx}&=&{\sinh(m\overline{r})\over m\sqrt{k +\sigma}},\\
\tan\left[m\sqrt{{k+\sigma}\over\sigma}(t - \overline{\tau})\right] &=& 
\sinh((m\overline{r}).
\end{eqnarray}
Straightforward calculation gives a new form of the metric (\ref{metS}):
\begin{eqnarray}
ds^2&=&d\overline{\tau}^2 - d\overline{r}^2 - {[k\,\sinh(m\overline{r}) + 
\sigma]\over m^2(k+\sigma)}\,d\overline{\varphi}^2 - dz^2 +{2\over m}
\sqrt{\sigma\over k+\sigma}\sinh(m\overline{r})\,d\overline{\tau}\,
d\overline{\varphi}\\
&=&\left(d\overline{\tau} + {1\over m}\sqrt{\sigma\over k+\sigma}
\sinh(m\overline{r})\,d\overline{\varphi}\right)^2 - \left(d\overline{r}^2 + 
{\cosh^2(m\overline{r})\over m^2}d\overline{\varphi}^2 + dz^2\right).
\label{metS1}
\end{eqnarray}
We thus immediately find the {\it fourth Killing} vector, 
$\partial_{\overline{\varphi}}$. In the original coordinates it reads:
\begin{equation}
 \xi_{(3)} ={\partial\over\partial\overline{\varphi}}={1\over m}
{\partial\over\partial x} - y{\partial\over\partial y}.\label{iso4}
\end{equation}

Alternatively, let us consider another transformation
$$
(t,\,x,\,y)\longrightarrow (\tau,\,r,\,\varphi)
$$
given by the formulae:
\begin{eqnarray}
e^{mx}&=&\cosh(mr) + \cos\varphi\,\sinh(mr),\label{tra1}\\
ye^{mx}&=&{\sin\varphi\,\sinh(mr)\over m\sqrt{k +\sigma}},\label{tra2}\\
\tan\left[{m\over 2}\sqrt{{k+\sigma}\over\sigma}(t - \tau)\right] &=& 
{\sin\varphi\over {\cos\varphi + \coth\left({mr\over 2}\right)}}.\label{tra3}
\end{eqnarray}
The metric (\ref{metS}) in the new coordinates reads:
\begin{eqnarray}
ds^2&=&d\tau^2 - dr^2 - {4\,\sinh^2\left({mr\over 2}\right)
[k\,\cosh^2\left({mr\over 2}\right) + \sigma]\over m^2(k+\sigma)}
\,d\varphi^2 - dz^2 \label{ctc} \\ &&\qquad\qquad\qquad\qquad\qquad
\qquad\qquad  + \,{4\over m}\sqrt{\sigma\over k+\sigma}\sinh^2\left(
\hbox{$\scriptstyle{mr\over 2}$}\right)\,d\tau\,d\varphi\nonumber\\
&=&\left(d\tau - {2\over m}\sqrt{\sigma\over k+\sigma}\sinh^2\left(
\hbox{$\scriptstyle{mr\over 2}$}\right)\,d\varphi\right)^2 - 
\left(dr^2 + {\sinh^2(mr)\over m^2}d\varphi^2
+ dz^2\right).\label{metS2}
\end{eqnarray}
Evidently $\partial_\varphi$ is also the Killing vector field. In the
original coordinates it looks somewhat complicate and can be represented
as a linear combination of three vectors:
\begin{equation}
{\partial\over\partial\varphi} = {1\over m}\sqrt{\sigma\over k + \sigma}\,
\xi_{(0)} + {1\over 2m\sqrt{k + \sigma}}\,\xi_{(1)} - 
\sqrt{k + \sigma}\,\xi_{(4)},
\end{equation}
where we denoted the {\it fifth Killing} vector:
\begin{equation}
 \xi_{(4)} = {\sqrt{\sigma}e^{-mx}
\over m(k+\sigma)}{\partial\over\partial t} + y{\partial\over\partial x}
+ {1\over 2}\left[{e^{-2mx}\over m(k+\sigma)} - 
my^2\right]{\partial\over\partial y}.\label{iso5}
\end{equation}

Three of the five Killing vectors (\ref{iso0}), (\ref{iso4}), (\ref{iso5}) 
satisfy commutation relations
 \begin{equation}
 [\xi_{(3)},\xi_{(1)}]=\xi_{(1)},\qquad
 [\xi_{(3)},\xi_{(4)}]=-\,\xi_{(4)},\qquad
 [\xi_{(1)},\xi_{(4)}]=m\xi_{(3)}.\label{comm01}
 \end{equation}
Thus $\xi_{(1)}, \xi_{(3)}, \xi_{(4)}$ form the closed subalgebra isomorphic 
to $so(2,1)$. Two Killing vectors $\xi_{(0)}, \xi_{(2)}$ span the
center of the isometry algebra, commuting with all its elements. 

In order to get some insight into possible causal problems, it is helpful
to compute squares of the Killing vectors. Straightforwardly we obtain:
\begin{eqnarray}
(\xi_{(0)}\cdot\xi_{(0)})&=&1,\\
(\xi_{(1)}\cdot\xi_{(1)})&=&-\,k\,e^{2mx},\\
(\xi_{(2)}\cdot\xi_{(2)})&=&-\,1,\\
(\xi_{(3)}\cdot\xi_{(3)})&=&-\left(k\,y^2\,e^{2mx} + {1\over m^2}\right),\\
(\xi_{(4)}\cdot\xi_{(4)})&=&-\,{k\over 4}\left(e^{mx}\,m\,y^2 + 
{e^{-mx}\over m(k + \sigma)}\right)^2,
\end{eqnarray}
The vector field $\xi_{(0)}$ is always timelike, its integral curves
coincide with the lines of coordinate time. The cases $k>0$, $k<0$ and
$k=0$ are essentially different, and it is reasonable to consider them
separately. 

For {\it positive} values of $k$ the four Killing vector fields $\xi_{(1)},
...,\xi_{(4)}$ are strictly spacelike; the first three of them provide 
spatial homogeneity of the $t=const$ hypersurfaces. We will demonstrate
the complete causality of the models with $k > 0$ later. In particular,
the closed timelike curves are absent in these geometries. 

For {\it negative} $k$, the vectors $\xi_{(1)}$ and $\xi_{(4)}$ become 
timelike, whereas $\xi_{(3)}$ can be both spacelike and timelike, depending 
on the values of spatial coordinates $x,y$. One can immediately verify the
existence of timelike closed curves, which was first noticed already
by G\"odel \cite{goe1}. Indeed, consider the closed curve $\{t(\varphi),
x(\varphi),y(\varphi)\}$, with $0\leq\varphi \leq 2\pi$, given by the 
coordinate transformation (\ref{tra1})-(\ref{tra3}) for $\tau=const, 
r=const, z=const$. From (\ref{ctc}) we find that for large enough $r$,
when $\cosh^2({mr\over 2}) > |{\sigma\over k}|$, such a curve is timelike. 
By construction, it is closed, and this may lead to the violation of
causality. Although in \cite{wright2} it was shown that such curves are
not geodesics, one can in principle imagine a motion of a physical 
particle along these curves under the action of certain forces. It is
worthwhile to note that the form of the metric is similar in two
different coordinate systems, compare (\ref{metS1}) and (\ref{metS2}).
However, for $k < 0$ the coordinate $\overline{\varphi}$ curve (with 
$\overline{\tau}=const, \overline{r}=const, z=const$) is also timelike for 
$\cosh^2(mr) > |{\sigma\over k}|$, but it is {\it not closed}. 

The {\it special case} $k=0$ was studied in detail by Rebou\c{c}as and 
Tiomno \cite{reb5}, see also \cite{reb9} and more recently \cite{roo}. 
The Riemannian curvature is simplified greatly for $k=0$. In particular,
the Weyl tensor (\ref{weyl0}) vanishes, and thus four-dimensional metric 
is conformally flat. At the same time, (\ref{gedcurv}) reduces to 
$R^{ab} = \omega^2\,\vartheta^a\wedge\vartheta^b$, with $a,b=0,1,2$.
The last relation describes a three-dimensional manifold of constant
curvature, i.e. (anti) de Sitter spacetime. This space has much wider
isometry group than $k\neq 0$ case. In addition to the 5 Killing vectors 
(\ref{iso0}), (\ref{iso4}), (\ref{iso5}), we easily find 2 more: 
\begin{eqnarray}
\xi_{(5)}&=& \sin(mt){\partial\over\partial t} - \cos(mt)
{\partial\over\partial x} + {e^{-mx}\over\sqrt{\sigma}}\sin(mt)
{\partial\over\partial y},\label{iso6}\\
\xi_{(6)}&=& \cos(mt){\partial\over\partial t} + \sin(mt)
{\partial\over\partial x} + {e^{-mx}\over\sqrt{\sigma}}\cos(mt)
{\partial\over\partial y}.\label{iso7}
\end{eqnarray}
Together with $\xi_{0}$ these vector fields form the second closed
subalgebra isomorphic to $so(2,1)$:
\begin{equation}
[\xi_{(0)},\xi_{(5)}]=m\xi_{(6)},\qquad
[\xi_{(0)},\xi_{(6)}]=-m\,\xi_{(5)},\qquad
[\xi_{(5)},\xi_{(6)}]=-m\xi_{(0)}.\label{comm02}
\end{equation}
The six Killing vectors $\xi_{(A)}$, $A=1,3,4,0,5,6$ with the commutation
relations (\ref{comm01}), (\ref{comm02}) describe the complete $so(2,2)$ 
algebra of isometries (algebra of the conformal group of a three-dimensional 
(anti) de Sitter spacetime) \cite{roo}. It is worthwhile to note that both 
new Killing vectors are spacelike:
\begin{equation}
(\xi_{(5)}\cdot\xi_{(5)})=(\xi_{(6)}\cdot\xi_{(6)})=-1,
\end{equation}

Starting from the metric in the form (\ref{metS2}) with $k=0$ inserted, 
we can make a simple coordinate transformation
$$
(\tau,\,r,\,\varphi)\longrightarrow (\tau,\,r,\,\theta),
$$
redefining the angular coordinate,
\begin{equation}
\varphi = {m\over 2}\,(\theta - \tau),
\end{equation}
which brings the metric to an explicitly (anti) de Sitter form:
\begin{equation}
ds^2 = \cosh^2\left(\hbox{$\scriptstyle{mr\over 2}$}\right)\,d\tau^2 -
dr^2 - \sinh^2\left(\hbox{$\scriptstyle{mr\over 2}$}\right)\,d\theta^2
- dz^2.\label{metS3}
\end{equation}
Note that $\omega={m\over 2}$ for $k=0$.

\section{Class of shear-free cosmological models with rotation and expansion}

 In \cite{banach,jetp1} we considered a wide class of viable cosmological 
 models with expansion and rotation. Let us describe it here briefly. 
 Denoting $x^0 = t$ as the cosmological time and $x^i , i=1,2,3$ as three
 spatial coordinates, we write the space-time interval in the form
 \begin{equation}
 ds^2 = dt^2 -2\,a\,n_{i}dx^{i}dt -a^2\,\gamma_{ij}\,dx^{i}dx^{j},\label{met0}
 \end{equation}
 where $a=a(t)$ is the scale factor, and
 \begin{equation}
 n_{i}=\nu_a\,e_i^{(a)},\quad \gamma_{ij}=\beta_{ab}\,e_i^{(a)}e_j^{(b)}.
 \label{ng}
 \end{equation}
 Here ($a,b=1,2,3$) $\nu_{a},\beta_{ab}$ are constant coefficients, while
 \begin{equation}
 e^{(a)}= e_{i}^{(a)}(x)\,dx^i \label{ea}
 \end{equation}
 are the invariant $1$--forms with respect to the action of a three-parameter
 group of motion which is admitted by the space-time (\ref{met0}). We assume
 that this group acts simply-transitively on the spatial ($t=const$) 
 hypersurfaces. It is well known that there exist 9 types of such manifolds,
 classified according to the Killing vectors $\xi_{(a)}$ and their commutators
 $[\xi_{(a)},\xi_{(b)}]=C^{c}{}_{ab}\xi_{(c)}$. Invariant forms (\ref{ea})
 solve the Lie equations ${\cal L}_{\xi_{(b)}}e^{(a)}=0$ for each Bianchi
 type, so that models (\ref{met0}) are spatially homogeneous. 

We consider $(t, x^i)$ as comoving coordinates for the cosmological matter,
that is the four-vector of average velocity of material fluid is equal
$u^\mu=\delta^\mu_0$. It is worthwhile to note that $u^\mu$ is not 
orthogonal to hypersurfaces of homogeneity. Such a models are usually called
``tilted". The kinematical characteristics of (\ref{met0}) are 
as follows: volume expansion is 
 \begin{equation}
 \vartheta = 3\,{\dot{a}\over a},\label{expan}
 \end{equation}
 nontrivial components of vorticity tensor are
\begin{equation}
\omega_{ij}=-\,{a\over 2}\,\hat{C}^{k}{}_{ij}\,n_{k},
\qquad i,j=1,2,3,\label{rot}
\end{equation}
 and shear tensor is trivial,
 \begin{equation}
 \sigma_{\mu\nu}=0.\label{shear}
 \end{equation}
 Hereafter the dot ($\,\dot{}\,$) denotes derivative with respect to the 
 cosmological time coordinate $t$. Tensor $\hat{C}^{k}{}_{ij}=
 e^{k}_{(a)}(\partial_{i}e^{(a)}_{j}-\partial_{j}e^{(a)}_{i})$ is the 
 anholonomity object for the triad (\ref{ea}); for I-VII Bianchi types values
 of its components numerically coincide with the corresponding structure 
 constants $C^{a}{}_{bc}$. The list of explicit expressions for $\xi_{(a)},
 e^{(a)}, C^{a}{}_{bc}, \hat{C}^{k}{}_{ij}$ for any Bianchi type is given in 
 \cite{banach}.

For completeness, let us mention that cosmological models (\ref{met0})
have nontrivial acceleration 
\begin{equation}
u^\nu\nabla_\nu u_\mu = (\,0, \,\dot{a}\,n_i\,).\label{acc}
\end{equation}

We choose the constant matrix $\beta_{ab}$ in (\ref{ng}) to be {\it positive
definite}. This important condition generalizes results of Maitra 
\cite{maitra}, and ensures the absence of closed time-like curves.

 One can immediately see that space-times (\ref{met0}) admit, besides tree
 Killing vector fields $\xi_{(a)}$, a nontrivial {\it conformal} Killing
 vector
 \begin{equation}
 \xi_{\rm conf}=a\,\partial_t .\label{conf}
 \end{equation}

 All models in the class (\ref{met0}) have a number of common remarkable
 properties \cite{banach,jetp1}:
 \begin{description}
 \item[Causality:]
 Space-time manifolds (\ref{met0}) are completely causal.

Indeed, following \cite{maitra}, let us suppose the spacetime (\ref{met0})
contains a closed curve $x^\mu(s)$ where $s$ is some parameter, $0\leq s
\leq 1$, which is timelike. The latter means that at any point on the curve
the length of the velocity vector is strictly positive,
\begin{equation}
g_{\mu\nu}\,{dx^\mu\over ds}{dx^\nu\over ds}\,>\,0.\label{timelike}
\end{equation}
Since the curve is closed, $x^\mu(0)=x^\mu(1)$, there exists a value $s_0$
at which ${dt\over ds}=0$. At this point one has
\begin{equation}
g_{\mu\nu}\,{dx^\mu\over ds}{dx^\nu\over ds}\Bigg\vert_{s=s_0} =
-\,a^2\,\beta_{ab}\,e^{(a)}_i e^{(b)}_j\,{dx^i\over ds}{dx^j\over ds},
\end{equation}
which is negative when $\beta_{ab}$ is positive definite. This contradicts
(\ref{timelike}) and hence we conclude that our assumption was wrong and
there are no closed timelike curves. 

For example, the stationary G\"odel type metric (\ref{metS}) represents
a particular case of (\ref{met0}) with $a=1$ and
\begin{equation}
\nu_a = (0, \sqrt{\sigma}, 0),\qquad 
\beta_{ab} =\left(\begin{array}{ccc}1 & 0 & 0\\ 0 & k & 0\\ 0 & 0 & 1
\end{array}\right),\qquad e^{(a)}_i=\left(\begin{array}{ccc}
1 & 0 & 0\\ 0 & e^{mx} & 0\\ 0 & 0 & 1 \end{array}\right).\label{betag}
\end{equation}
It is clear that $\beta_{ab}$ in (\ref{betag}) is positive definite for $k>0$. 
Negative values of $k$, as we saw explicitly in Sec.~\ref{statged}, allow for 
the closed timelike curves. In particular, this is true for the original 
G\"odel solution (\ref{gedpar}). 

 \item[Isotropy of MBR:]
The Microwave Background Radiation (MBR) is totally isotropic in (\ref{met0})
for any moment of the cosmological time $t$.

In the geometric optics approximation, light propagates along null geodesics,
i.e. curves $x^\mu(\lambda)$ with certain affine parameter $\lambda$ and
tangent vector $k^\mu=dx^\mu/d\lambda$ which satisfy
\begin{equation}
k^\nu\nabla_\nu k^\mu =0,\qquad k^\mu\,k_\mu =0.\label{light}
\end{equation}
As it is well known, see e.g. \cite{ellis1,ellis2}, the red shift $Z$, which
reflects the dependence of frequency on the relative motion of source and 
observer, is given by the simple formula
\begin{equation}
1 + Z = {(k^\mu u_\mu)_S\over (k^\mu u_\mu)_P},\label{red}
\end{equation}
where the subscripts ``{\it S}" and ``{\it P}" refer, respectively, to the
spacetime points where a source emitted a ray and an observer detected it.

MBR is characterized by the black-body spectrum. One can show \cite{ellis1}
that the temperature of the black-body radiation depends on the red shift,
\begin{equation}
T_P = {T_S\over 1 + Z} = T_S\,{(k^\mu u_\mu)_P\over (k^\mu u_\mu)_S}.
\label{Toe}
\end{equation}
In general, as was noticed, e.g., in \cite{haw,colhaw,temp1,bjs,temp2},
the observed temperature depends on the direction of observation through
the ratio of $(k^\mu u_\mu)$'s. However, it is not the case for the class
of shear-free models (\ref{met0}). Indeed, from (\ref{light}) we easily 
prove that the scalar product of the wave vector with any Killing vector
or conformal Killing vector is constant along the ray. In particular, for
the conformal Killing (\ref{conf}) we immediately obtain the first integral
of the null-geodesics equations (\ref{light}):
\begin{equation}
k_\mu\,\xi^\mu_{\rm conf} = {\rm const}\qquad \Longrightarrow\qquad
a(t_P)(u^\mu k_\mu)_P =a(t_S)(u^\mu k_\mu)_S,\label{KU}
\end{equation}
where we used the evident proportionality of the Killing vector to the
average four-velocity of matter,
\begin{equation}
\xi^\mu_{\rm conf} = a(t)\,u^\mu.\label{confU}
\end{equation}
Substituting (\ref{KU}) into (\ref{Toe}), one finds
\begin{equation}
T_P = {T_S\over 1 + Z} = T_S\,{a(t_S)\over a(t_P)}.
\end{equation}
Thus, in the class of cosmological models (\ref{met0}), MBR is completely
isotropic, exactly like in the standard cosmology. As a consequence, the
earlier strong restrictions of the global rotation derived from the isotropy
of MBR \cite{haw,colhaw,temp1,bjs,temp2}, are not applicable to the class of
models under consideration. At the same time, the result obtained by no means
contradicts the earlier observations, all of which were made for cosmologies
{\it with shear}. Here we have trivial shear (\ref{shear}), and thus one
clearly concludes that it is shear, not vorticity, which is the true source
of anisotropy of MBR. 

\item[No parallaxes:]
Rotation (\ref{rot}) does not produce parallax effects.

In a very interesting paper \cite{hasse2}, Hasse and Perlick introduced the
generalized notion of parallax in cosmology. Let $S_1$ and $S_2$ be any two
sources, seen by an observer $P$. Provided all the three objects do not have
peculiar motion  with respect to the matter with the average four-velocity
$u^\mu$, one says of a parallax effect when the angle between the rays coming
from $S_1$ and $S_2$ to $P$, changes in time. Six equivalent conditions were
formulated in \cite{hasse2} which are sufficient for the absence of parallax.
One can easily check that all of them are satisfied for the class of models
(\ref{met0}). For example, according to the P(3) condition (in notation of
\cite{hasse2}), the velocity $u^\mu$ should be proportional to some conformal
Killing vector. Indeed, this is the case, (\ref{confU}). Alternatively,
according to P(4) there should exist a function $f$ such that 
$$
{1\over 2}{\cal L}_u \,g_{\mu\nu} = (u^\lambda\partial_\lambda f)\,g_{\mu\nu} 
- u_{(\mu}\partial_{\nu)}f.
$$
One verifies straightforwardly that $f=\ln a$. According to P(5), $u^\mu$ is
shear-free and the one-form 
$$
\rho:=\left(u^\nu\nabla_\nu u_\mu - {\vartheta\over 3}\,u_\mu\right)dx^\mu
$$
is closed, $d\rho =0$. From (\ref{acc}) and (\ref{expan}), we find $\rho=0$,
hence P(5) is trivially fulfilled. Finally, according to P(6), there should
exist a ``red shift potential". One can prove that again the latter if given
by the function $f=\ln a$. 

We thus conclude, that the models (\ref{met0}) are parallax-free, and 
consequently, the value of vorticity cannot be determined from parallax
effects, contrary to the claims of \cite{rub,tre1,tre2}.
 \end{description}

Summarizing, cosmological models with rotation and expansion (\ref{met0})
solve the first three problems of cosmic rotation, and the most strong 
limits on the cosmic rotation, obtained earlier from the study of MBR 
\cite{haw,colhaw,temp1} and of the parallaxes in rotating world 
\cite{tre1,rub}, are not true for this class of cosmologies. This class of 
metrics is rich enough, as it contains all kinds of worlds: open and closed 
ones with different topologies.  

\section{G\"odel type expanding cosmological model}

To the end of this paper we will consider now the natural non-stationary
generalization of the original G\"odel metric (\ref{gd}) which has drawn
considerable attention in the literature. This generalized model is obtained
from (\ref{metS}) by introducing the time-dependent scale factor $a(t)$,
\begin{equation}
ds^2 = dt^2 - 2\sqrt{\sigma}a(t)e^{mx}dtdy -
a^2 (t)(dx^2 + ke^{2mx}dy^2 + dz^2).\label{met1}
\end{equation}
Here, $x^1 =x, x^2 =y, x^3 = z$, and the metric (\ref{met1}) evidently
belongs to the family (\ref{met0}) with (\ref{betag}). As we saw already,
the condition $k > 0$ guarantees the absence of closed time-like curves. 
The metric (\ref{met1}) is usually called the {\it G\"odel type model} with 
rotation and expansion. Coordinate $z$ gives the direction of the global 
rotation, the magnitude of which
\begin{equation}
\omega = \sqrt{{1\over 2}\omega_{\mu\nu}\omega^{\mu\nu}}=
{m\over 2a}\sqrt{\sigma\over {k+\sigma}}\label{rotged}
\end{equation}
decreases in expanding world [compare with (\ref{rotged0})].

The three Killing vector fields are, recall (\ref{iso0}) and (\ref{iso4}),
\begin{equation}
\xi_{(1)} =\partial_y ,\qquad  
\xi_{(2)} =\partial_z ,\qquad
\xi_{(3)} ={1\over m}\partial_x - y\partial_y ,\quad  .\label{kil}
\end{equation}
These satisfy commutation relations
\begin{equation}
[\xi_{(1)},\xi_{(2)}]=\xi_{(2)},\qquad
[\xi_{(1)},\xi_{(3)}]=[\xi_{(2)},\xi_{(3)}]=0,\label{comm}
\end{equation}
showing that the model (\ref{met1}) belongs to the Bianchi type III.

As before, it is convenient to choose at any point of the spacetime 
(\ref{met1}) a local orthonormal (Lorentz) tetrad $h^{a}_{\mu}$. For 
simplicity, we will work with the immediate replacement of (\ref{tetrad0}):
\begin{equation}
h^{\hat{0}}_{0}=1,\quad h^{\hat{0}}_{2}=-a\sqrt{\sigma}e^{mx},\quad
h^{\hat{1}}_{1}=h^{\hat{3}}_{3}=a,\quad h^{\hat{2}}_{2}=ae^{mx}
\sqrt{k+\sigma}. \label{tetrad}
\end{equation}

From the point of view of the Petrov classification, one can verify that
the G\"odel type model is of the type $D$. Indeed, a straightforward 
calculation yields an obvious generalization of (\ref{weyl0}):
\begin{equation}
C_{\hat{0}\hat{1}\hat{0}\hat{1}} = C_{\hat{0}\hat{2}\hat{0}\hat{2}} =
-\,C_{\hat{2}\hat{3}\hat{2}\hat{3}} = -\,C_{\hat{3}\hat{1}\hat{3}\hat{1}}
={1\over 2}\,C_{\hat{1}\hat{2}\hat{1}\hat{2}} = 
-\,{1\over 2}\,C_{\hat{0}\hat{3}\hat{0}\hat{3}} = {m^2\over 6a^2(t)}
\left({k\over k + \sigma}\right).\label{weyl1}
\end{equation}

\section{Dynamical realizations}

Several dynamical realizations (i.e. construction of exact cosmological
solutions for the gravitational field equations) of the shear-free models
(\ref{met0}), and in particular of the G\"odel type metric (\ref{met1}), 
are known. In the Einstein's general relativity theory the G\"odel type
models were described in 
\cite{krepan,5,panov1,panov2,panov3,panov4,panov5,panov6,panov7,panov8,panov9}, 
with different matter sources, whereas the Bianchi type II and III solutions 
(\ref{met0}) were obtained in \cite{reb2,n1} for the imperfect fluid with 
heat flow and cosmological constant. 

In this section, we will describe a model \cite{io} in which the gravitational 
field dynamics is determined by the minimal quadratic Poincar\'e gauge model. 
which is the closest extension of the Einstein's general relativity theory. 

\subsection{Field equations}

In the framework of Poincar\'e gauge theory, gravitational field is 
described by the tetrad $h^{a}_{\mu }$ and the local Lorentz connection
${\Gamma }^{a}{}_{b\mu }$. In general case, gravitational Lagrangian  
is constructed as an invariant contraction from the curvature tensor 
\begin{equation}
{R}^{a}{}_{b\mu\nu}= \partial_{\mu}{\Gamma}^{a}{}_{b\nu}
- \partial _{\nu }{\Gamma}^{a}{}_{b\mu } + 
{\Gamma}^{a}{}_{c\mu}{\Gamma }^{c}{}_{b\nu } - 
{\Gamma}^{a}{}_{c\nu}{\Gamma }^{c}{}_{b\mu },
\end{equation}
and the torsion tensor
\begin{equation}
T^{a}{}_{\mu \nu }= \partial _{\mu }h^{a}_{\nu } -  
\partial _{\nu }h^{a}_{\mu } + \widetilde{\Gamma }^{a}{}_{b\mu }h^{b}_{\nu } - 
\widetilde{\Gamma }^{a}{}_{b\nu }h^{b}_{\mu }.
\end{equation}
Note that here our notations for the connection, curvature and torsion are
slightly different from \cite{fwh4}. In order to illustrate some common 
properties of cosmological models with rotation and expansion, here we 
consider the so called minimal quadratic Poincar\'e gauge gravity \cite{io} 
which is the closest extension of Einstein's general relativity (we may also 
call such a model the generalized Einstein-Cartan theory). The corresponding 
Lagrangian reads:
\begin{equation}
L_{g}= - {1\over 16\pi G}\left[{R} + b\,{R}^{2} 
+ a_{1}T^{\alpha}{}_{\mu \nu }T_{\alpha}{}^{\mu \nu } + 
a_{2}T_{\alpha \mu \nu }T^{\mu \alpha \nu } + 
a_{3}T_{\mu }T^{\mu }\right].\label{1}
\end{equation}
Here ${R} = h^{\mu }_{a}h^{\nu b}{R}^{a}{}_{b\mu \nu }$ 
-- the Riemann-Cartan curvature scalar, $T^\alpha{}_{\mu\nu}=h^\alpha_a
T^a{}_{\mu\nu}$, and $T_{\mu}= T^{\lambda }{}_{\mu \lambda }$
-- is the torsion trace,  $b, a_{1}, a_{2}, a_{3}$ are the coupling
constants. 

Independent variation of the action (\ref{1}) with respect to $h^a_\mu$ 
and $\widetilde{\Gamma}^{a}{}_{b\mu}$ yields the field equations:
\begin{eqnarray} 
&&{R}_{\mu \nu } - {1\over 2}g_{\mu\nu}{R} + 2b{R}\left({R}_{\mu \nu } - 
{1\over 4}g_{\mu\nu}{R}\right) \nonumber\\
&+& a_{1}\left( -\,{1\over 2}\,g_{\mu \nu }T_{\alpha \beta \gamma }
T^{\alpha\beta\gamma} + 2T^{\alpha}{}_{\mu\beta}T_{\alpha\nu}{}^{\beta} - 
2T_{\alpha }T_{\nu \mu }{}^{\alpha } -\, 2{\nabla }_{\alpha }
T_{\nu \mu }{}^{\alpha } - T_{\mu\alpha\beta}T_{\nu }{}^{\alpha\beta}\right)
\nonumber\\
&+& a_{2}\left( - \,{1\over 2}\,g_{\mu\nu}T_{\alpha\beta\gamma}
T^{\beta\alpha\gamma} + T^{\alpha }{}_{\beta\mu}
T^{\beta}{}_{\alpha\nu} + T_{\alpha}T^{\alpha}{}_{\nu\mu} - 
T_{\alpha}T_{\mu\nu}{}^{\alpha} - {\nabla }_{\alpha }T_{\mu \nu }{}^{\alpha } 
+ {\nabla }_{\alpha }T^{\alpha }{}_{\nu \mu }\right) 
\nonumber\\
&+& a_3\left( {1\over 2}\,g_{\mu \nu }T_{\alpha }T^{\alpha } 
- {\nabla }_{\nu }T_{\mu } + g_{\mu \nu }{\nabla}_\alpha
T^{\alpha } \right) = \kappa\Sigma_{\mu\nu},\label{2}
\end{eqnarray}
\begin{eqnarray}
&-&\left({1 + 2b{R}\over 2}\right)\left(T^{\alpha }{}_{\mu \nu } + 
2\delta^{\alpha }_{[\mu }T_{\nu ]}\right) - 
2b\delta^{\alpha }_{[\mu }\partial_{\nu ]}{R} \nonumber\\
&+& \left(2a_1 - a_{2}\right)T_{[\mu \nu]}{}^{\alpha } - {a_2} 
T^{\alpha }{}_{\mu \nu } + {a_3}\delta^{\alpha }_{[\mu}T_{\nu]}
= \kappa \tau^{\alpha}{}_{\mu\nu},\label{3}
\end{eqnarray}
where $\kappa = 8\pi G$, with $\tau^{\alpha}{}_{\mu\nu}$ and $\Sigma_{\mu\nu}$ 
as the canonical tensors of spin and energy-momentum, respectively.

Equation (\ref{3}) generalizes the well known algebraic relation between
torsion and spin in the Einstein-Cartan theory. The tensors of torsion
and spin may be decomposed into irreducible parts: trace $T_{\mu }$ and 
$\tau_{\mu }= \tau^{\lambda }{}_{\mu \lambda }$, pseudotrace 
${\stackrel{\vee}{T}}{}_{\mu}={1\over 2}\epsilon_{\mu\nu\alpha\beta}
T^{\nu\alpha\beta}$ and ${\stackrel{\vee}{\tau}}{}_{\mu}= {1\over 2}
\epsilon_{\mu\nu\alpha\beta}\tau^{\nu\alpha\beta}$, and trace-free parts 
$\overline{T}^{\alpha}{}_{\mu\nu}$ and $\overline{\tau}^{\alpha}{}_{\mu\nu}$:
\begin{eqnarray}
T^{\alpha }{}_{\mu\nu} &=& \overline{T}^{\alpha }{}_{\mu \nu } + 
{2\over 3}T_{[\mu}\delta ^{\alpha }_{\nu ]} + {1\over 3}\epsilon^{\alpha}
{}_{\mu\nu\lambda}{\stackrel{\vee}{T}}{}^{\lambda},\label{4.1}\\
\tau^{\alpha}{}_{\mu\nu} &=& \overline{\tau}^{\alpha }{}_{\mu \nu } + 
{2\over 3}\tau_{[\mu }\delta^{\alpha }_{\nu ]} + {1\over 3}\epsilon^{\alpha }
{}_{\mu \nu \lambda }{\stackrel{\vee}{\tau}}{}^{\lambda }.\label{4.2}
\end{eqnarray}
Easy to see that eq. (\ref{3}) decomposes into three equations for
irreducible components:
\begin{eqnarray}
\left(\mu_{2} + 2b{R}\right)T_{\mu} + 3b\,\partial_{\mu }{R} &=& 
\kappa \tau_{\mu },\label{5}\\
\left(\mu_{1}- bR\right){\stackrel{\vee}{T}}{}_{\mu} &=& \kappa 
{\stackrel{\vee}{\tau}}{}_{\mu},\label{6}\\
\left(\mu_{3} - bR\right)\overline{T}^{\alpha }{}_{\mu\nu} &=& 
\kappa\overline{\tau}^{\alpha }{}_{\mu\nu},\label{7}
\end{eqnarray}
where we denoted the following combinations of the coupling constants
\begin{eqnarray}
\mu _{1} &=& 2a_{1} - 2a_{2} - {1\over 2},\label{8.1}\\
\mu _{2} &=& 1 - {1\over 2}\left(2a_{1} + a_{2} + 3a_{3}\right),\label{8.2}\\
\mu _{3} &=& -\left({1\over 2} + a_1 + {a_2\over 2}\right),\label{8.3}
\end{eqnarray}
which play important role in the determination of mass for quanta of
torsion and couplings in all spin sectors of quadratic Poincar\'e gauge
gravity model.

The eq. (\ref{5}) shows that in contrast to the other irreducible parts,
the torsion trace is characterized by the differential and not merely
algebraic coupling. The Riemann-Cartan curvature scalar plays a role of
its ``potential".  Contraction of (\ref{2}) yields an equation for ${R}$,
\begin{equation}
- {R} + {1\over 2}\left(1 + 2\mu_{3}\right)\overline{T}_{\alpha\mu\nu}
\overline{T}^{\alpha\mu\nu} + {1\over 6}\left(1 + 2\mu_{1}\right)
{\stackrel{\vee}{T}}{}_{\mu}{\stackrel{\vee}{T}}{}^{\mu} +
2\left(1 - \mu_{2}\right)\left(\widetilde{\nabla}_{\mu }T^{\mu } -
{1\over 3}T_{\mu }T^{\mu }\right) = \kappa \Sigma, \label{9}
\end{equation}
where $\Sigma = g_{\mu\nu}\Sigma^{\mu\nu}$ is the trace of the
energy-momentum tensor. Inserting (\ref{5})-(\ref{7}) into (\ref{9}), 
we find a non-linear differential equation for the curvature scalar.
As usual, hereafter we denote by tilde the purely Riemannian (torsion-free)
geometrical objects and operators. 

In view of specific nature of the torsion trace, it is convenient to
separate its contribution to the eq. (\ref{2}). Let us introduce
\begin{equation}
\widehat{T}^{\alpha}{}_{\mu \nu } = \overline{T}^{a}{}_{\mu \nu } + {1\over 3}
\epsilon^{\alpha}{}_{\mu\nu\lambda}{\stackrel{\vee}{T}}{}^{\lambda},\label{10}
\end{equation}
and hereafter hat ``$\,\widehat{}\,$" will denote the corresponding
geometrical quantities constructed with the help of the Riemann-Cartan
connection with the trace-free torsion (\ref{10}). The symmetric 
part of (\ref{2}) then is rewritten as
\begin{eqnarray}
(1 &+& 2b{R})\left(\widehat{R}_{(\mu\nu)} - {1\over 2}
g_{\mu\nu}\widehat{R}\right) + {1\over 2}b{R}^{2}g_{\mu\nu} 
+ \left({{1+2\mu_3}\over 12}\right)T^{\lambda}\widehat{T}_{(\mu\nu)\lambda} 
\nonumber\\
&+& a_{1}\left( - \,{1\over 2}\,g_{\mu\nu}\widehat{T}_{\alpha\beta\gamma}
\widehat{T}^{\alpha\beta\gamma} + 2\widehat{T}^{\alpha}{}_{\mu\beta}
\widehat{T}_{\alpha\nu}{}^{\beta} -
\widehat{T}_{\mu\alpha\beta}\widehat{T}_{\nu}{}^{\alpha\beta} -
2\widehat{\nabla}_{\alpha}\widehat{T}_{(\mu\nu)}{}^{\alpha}\right) 
\nonumber\\
&+& a_{2}\left( - \,{1\over 2}\,g_{\mu\nu}\widehat{T}_{\alpha\beta\gamma}
\widehat{T}^{\beta\alpha\gamma} + \widehat{T}_{\alpha\beta\mu}
\widehat{T}^{\beta\alpha}{}_{\nu} - \widehat{\nabla}_{\alpha}
\widehat{T}_{(\mu\nu)}{}^{\alpha}\right) 
\nonumber\\
&+& \left({{8b{R} + 4\mu_2}\over 3}\right)\left({\nabla}_{(\mu}T_{\nu)} 
- g_{\mu\nu}{\nabla}_{\alpha}T^{\alpha} - {1\over 6}g_{\mu\nu}
T_{\alpha}T^{\alpha}\right) = \kappa\Sigma_{(\mu\nu)} ,\label{11}
\end{eqnarray}
whereas the antisymmetric part of (\ref{2}) is an identity in view of
the geometric relation
\begin{equation}
2{R}_{[\mu\nu]} + \left({\nabla}_{\alpha} + T_{\alpha }\right)
\left(T^{\alpha}{}_{\mu\nu} + 2\delta^{\alpha}_{[\mu}T_{\nu]}\right) = 0,
\end{equation}
and the angular momentum conservation law
\begin{equation}
\left({\nabla}_{\alpha} + T_{\alpha}\right)
\tau^{\alpha}{}_{\mu\nu} = \Sigma_{[\mu\nu]} .
\end{equation}
Inserting the torsion ${\stackrel{\vee}{T}}{}_{\mu}$ and $\overline{T}^{\alpha}
{}_{\mu\nu}$ from (\ref{6}) and (\ref{7}) into (\ref{10}), and further into 
(\ref{11}), one can write, similarly to the Einstein-Cartan theory, the 
effective Einstein equations for the metric. Spin of matter will contribute 
to the so called modified energy-momentum tensor.

\subsection{Matter sources}

We will describe cosmological matter by means of phenomenological model
of a spinning fluid of Weyssenhoff and Raabe. This is a classical model 
of a continuous medium the elements of which are characterized, along with
energy and momentum, by an intrinsic angular momentum (spin). Usually, 
hydrodynamical approach is considered as a good approximation to the
description of realistic cosmological matter on the early, as well as on
the later stages of universe's evolution. It seems natural to take into
account spin (proper angular momentum) of elements of cosmological fluid,
i.e., particles on early stage and galaxies and clusters of galaxies 
on later stages. In general case, an element of the Weyssenhoff fluid
is characterized by the tensor of spin density $\tau_{\mu\nu}$, charge $e$ 
and (proportional to spin) magnetic moment $\chi\tau_{\mu\nu}$ ($\chi$ -- 
is a constant).  The consistent variational theory of spin fluid in the
Riemann-Cartan spacetime, developed in \cite{flu1}, yields the 
following currents in the right-hand sides of the gravitational field
equations (\ref{2}) and (\ref{3}):
\begin{eqnarray}
\Sigma_{\mu\nu} &=& - pg_{\mu\nu} + u_\mu\left[u_\nu\left(\epsilon  + p + 
\chi\tau_{\alpha\beta}F^{\alpha\beta}\right) + 2u^\alpha{\nabla}_{\beta}
\tau^{\beta}{}_{\alpha\nu}\right]\nonumber\\
&&- F_{\mu \alpha }F_{\nu }{}^{\alpha } + {1\over 4}g_{\mu \nu }
F_{\alpha \beta }F^{\alpha \beta } - 2\chi \left(\tau_{\mu \alpha }
\tau_{\nu }{}^{\alpha } + \tau_{\nu \beta }F^{\alpha \beta }u_{\alpha }
u_{\mu }\right),\label{12}\\
\tau^{\alpha}{}_{\mu\nu} &=& u^{\alpha}\tau_{\mu\nu},\label{13}
\end{eqnarray}
where $u^{\mu}$ is fluid's four-velocity, $p$ is pressure,  
$\epsilon$  is the internal energy density, $F_{\mu\nu} = \partial_{\mu}
A_{\nu} - \partial_{\nu}A_{\mu}$ is the electromagnetic field created by
and interacting with the electrodynamical characteristics of the medium.

Dynamics of the fluid is given by the rotational and translational
equations of motion:
\begin{eqnarray}
\left({\nabla}_{\alpha} + T_{\alpha}\right)
\tau^{\alpha}{}_{\mu\nu} &=& u_{\mu}u^{\lambda}{\nabla}_{\alpha}
\tau^{\alpha}{}_{\lambda\nu} - u_{\nu}u^{\lambda}
{\nabla}_{\alpha}\tau^{\alpha}{}_{\lambda\mu} \nonumber\\
&&+ \chi F^{\alpha}{}_{\beta}\left[\left(\delta^{\beta}_{\mu} - 
u^{\beta}u_{\mu}\right)\tau_{\alpha\nu} - \left(\delta^{\beta}_{\nu} - 
u^{\beta}u_{\nu}\right)\tau_{\alpha\mu}\right],\label{14}\\
\left({\nabla}_{\nu} + T_{\nu}\right)\Sigma^{\nu}{}_{\mu} 
&-& T^{\alpha}{}_{\mu\beta}\Sigma^{\beta }{}_{\alpha } - 
\tau^{\nu}{}_{\alpha\beta}{R}^{\alpha \beta }{}_{\mu \nu } = 0.\label{15}
\end{eqnarray}
The self-consistent variational principle \cite{flu1} yields also the
Maxwell's equations for the electromagnetic field and the conservation
law of number of particles:
\begin{eqnarray}
\widetilde{\nabla}_{\mu}F^{\mu\nu} &=&
e\rho u^{\nu} + 2\chi\widetilde{\nabla}_{\mu}\tau^{\mu\nu},\label{16}\\
\widetilde{\nabla}_{\mu}\left(\rho u^{\mu}\right) &=& 0,\label{17}
\end{eqnarray}
where $\rho$ is the particle density.

The system of differential and algebraic equations (\ref{2}),  (\ref{3}),  
(\ref{12})-(\ref{17}) is complete, and one can determine all the 
gravitational and material physical variables from this system. It is
worthwhile to remind that spin satisfies the Frenkel condition,
\begin{equation}
u^{\mu}\tau_{\mu\nu} = 0.\label{18}
\end{equation}
The specific features of dynamics of the Weyssenhoff spin fluid in the
Riemann-Cartan spacetime were studied in detail in \cite{flu1}.

\subsection{Exact solution for the G\"odel type universe}

Let us describe cosmological model with rotation and expansion which 
provides an exact solution of the gravitational field equations. We will
restrict ourselves to the shear-free class of spatially homogeneous metrics 
(\ref{met0}), and for definiteness, a particular case -- the G\"odel type
model (\ref{met1}) -- will be considered in detail. 

The field equations (\ref{2}) and (\ref{3}) with the sources (\ref{12}),
(\ref{13}) are most conveniently formulated with respect to (nonholonomic)
local Lorentz frame. Let us choose the coframe (\ref{tetrad}). Its inverse
reads:
\begin{equation}
h^{0}_{\hat{0}}= 1,\quad h^{0}_{\hat{2}}=\sqrt{\sigma\over{k+\sigma}},
\quad h^{3}_{\hat{3}}= h^{1}_{\hat{1}}= {1\over a} ,\quad 
h^{2}_{\hat{2}}= {1\over ae^{mx}\sqrt{k+\sigma}},\label{24.2}
\end{equation}
Direct calculation of the local Lorentz connection $\widetilde{\Gamma}^{a}
{}_{b\mu}= h^{a}_{\alpha}h^{\beta}_{b}\widetilde{\Gamma}^{\alpha}_{\beta\mu} 
+ h^a_\alpha\partial_\mu h^\alpha_b$ (with $\widetilde{\Gamma}
{}^\alpha_{\beta\mu}$ as the Christoffel symbols) yields for the metric
(\ref{met1}):
\begin{eqnarray}
\widetilde{\Gamma}^{\hat{0}}{}_{\hat{2}\hat{0}} &=& 
\widetilde{\Gamma}^{\hat{1}}{}_{\hat{2}\hat{1}} 
= \widetilde{\Gamma}^{\hat{3}}{}_{\hat{2}\hat{3}} =
{\dot{a}\over a}\,\sqrt{\sigma\over k +\sigma},\qquad 
\widetilde{\Gamma}^{\hat{1}}{}_{\hat{2}\hat{2}} = - \,{m\over a} , \\
\widetilde{\Gamma}^{\hat{2}}{}_{\hat{0}\hat{1}} &=& 
\widetilde{\Gamma}^{\hat{2}}{}_{\hat{1}\hat{0}} 
= -\,\widetilde{\Gamma}^{\hat{0}}{}_{\hat{1}\hat{2}} = 
{m\over 2a}\,\sqrt{\sigma\over k + \sigma}, \label{25}
\end{eqnarray}
where $\widetilde{\Gamma}^{a}{}_{bc} = 
\widetilde{\Gamma}^{a}{}_{b\mu}h^{\mu}_{c}$ .

Since the direction of rotation (along the $z$ axis) is apparently a
distinguished one, it is natural to assume that the spin of the fluid and
electromagnetic field are also oriented correspondingly along $z$. For
the medium with vanishing magnetic moment of its elements ($\chi =0$)
one finds from (\ref{14})-(\ref{18}):
\begin{eqnarray}
\tau_{\hat{1}\hat{2}} &=& \tau_0\,a^{-3}, \label{26}\\
F_{\hat{1}\hat{2}} &=& B_{0}\,a^{-2}, \label{27}\\
\rho &=& \rho_{0}\,a^{-3}, \label{28}
\end{eqnarray}
where the integration constants are related by the consistency condition
\begin{equation}
mB_{0}\,\sqrt{\sigma\over k + \sigma} = e\,\rho_{0}. \label{29}
\end{equation}
We will now describe a solution of the gravitational field equations for
which the Riemann-Cartan curvature scalar is assumed to be constant:
\begin{equation}
{R} = {\rm const}.  \label{30}
\end{equation}
Then in view of (\ref{18}), eqs. (\ref{5})-(\ref{7}) yield the irreducible
parts of torsion: 
\begin{equation}
T_{\mu}= 0,\qquad {\stackrel{\vee}{T}}{}_{\mu}= {\lambda_1\over 2}
\epsilon_{\mu\nu\alpha\beta}u^{\nu}\tau^{\alpha\beta},\qquad
\overline{T}^{\alpha}{}_{\mu\nu}= {\lambda_3\over 3}\left(2u^{\alpha}
\tau_{\mu\nu} + u_{\mu}\tau^{\alpha}{}_{\nu} - u_{\nu}\tau^{\alpha}{}_{\mu}
\right),\label{31}
\end{equation}
where we denoted the constants
\begin{equation}
\lambda_{1}= {\kappa \over \mu_{1} - bR},\qquad 
\lambda_{3}= {\kappa \over \mu_{3} - bR}.\label{32}
\end{equation}
Nontrivial solution exists only when $\mu_{1}\neq b{R}\neq\mu_{3}$,
otherwise (\ref{6}) and (\ref{7}) lead to the vanishing spin. As for the
eq. (\ref{5}),  one has two alternatives: if $\mu_{2} + 2b{R}\neq 0$, then 
the torsion trace is zero, whereas if $\mu _{2} + 2b{R} = 0,$ then $T_{\mu}$ 
is arbitrary. However in both cases the torsion trace completely decouples
from the eq. (\ref{11}), hence without restricting generality we can put
$T_{\mu}= 0$. 

Using (\ref{31}),  we can rewrite (\ref{10}) as
\begin{equation}
\widehat{T}^{\alpha}{}_{\mu\nu}= {1\over 3}\left[\left(2\lambda_{3}+
\lambda_{1}\right)u^{\alpha}\tau_{\mu\nu} + \left(\lambda_{3}-\lambda_{1}
\right) u_{\mu}\tau^{\alpha}{}_{\nu} + \left(\lambda_{1}-\lambda_{3}\right)
u_{\nu}\tau^{\alpha}{}_{\mu}\right].
\end{equation}
Substituting this expression into (\ref{11}), one finds
\begin{eqnarray}
&&a_{1}\left( - \,{1\over 2}\,g_{\mu\nu}\widehat{T}_{\alpha\beta\gamma}
\widehat{T}^{\alpha\beta\gamma} + 2\widehat{T}^{\alpha}{}_{\mu\beta}
\widehat{T}_{\alpha\nu}{}^{\beta} - \widehat{T}_{\mu\alpha\beta}
\widehat{T}_{\nu}{}^{\alpha\beta} - 2\widehat{\nabla}_{\alpha}
\widehat{T}_{(\mu\nu)}{}^{\alpha}\right) \nonumber\\ &&\quad 
+ a_{2}\left( - \,{1\over 2}\,g_{\mu\nu}\widehat{T}_{\alpha\beta\gamma}
\widehat{T}^{\beta\alpha\gamma} + \widehat{T}_{\alpha\beta\mu}
\widehat{T}^{\beta\alpha}{}_{\nu} - \widehat{\nabla}_{\alpha}
\widehat{T}_{(\mu\nu)}{}^{\alpha}\right) =\nonumber\\
&&= {1\over 18}\left(2\lambda_{3} + \lambda_{1}\right)\left[\lambda _{1}
\left(1 + 2\mu_{1}\right) - \lambda_{3}\left(1 + 2\mu_{3}\right)\right]
\tau_{\mu\alpha}\tau_{\nu}{}^{\alpha} \nonumber\\
&&\quad + \left({\lambda_1 - 4\lambda_3\over 36}\right)\left[\lambda_{1}
\left(1 + 2\mu_{1}\right) + 2\lambda_{3}\left(1 + 2\mu_{3}\right)\right]
u_{\mu}u_{\nu}\tau_{\alpha\beta}\tau^{\alpha\beta} \nonumber\\
&&\quad + {1\over 24}\left[4\lambda^{2}_{3}\left(1 + 2\mu_{3}\right) - 
\lambda^{2}_{1}\left(1 + 2\mu_{1}\right)\right]\,g_{\mu\nu}\,
\tau_{\alpha\beta}\tau^{\alpha\beta} + \lambda_{3}\left(1 + 2\mu_{3}\right)
\widetilde{\nabla}_{\alpha}\left[u_{(\mu}\tau_{\nu )}{}^{\alpha}\right].
\label{33}
\end{eqnarray}
After these computations, one finally finds that when the constant
in (\ref{30}) is equal ${R} = - 1/2b$, the equations (\ref{11}) are
reduced to the algebraic relations between the pressure $p$, energy
density $\epsilon$, spin $\tau_{\alpha\beta}$ and Maxwell tensor
$F_{\alpha\beta}$. From these equations, in addition to (\ref{29})
we find a new relation between the integration constants,
\begin{equation}
B^{2}_{0} = -\,m\,\tau_{0}\,\sqrt{\sigma\over k + \sigma}.\label{34}
\end{equation}
Without restricting generality, one can choose the parameter $m$ to be 
positive, then the minus sign in (\ref{34}) means that the spin and 
vorticity have the same direction. Thus $\tau_{0}<0$ [from (\ref{rot}), 
(\ref{tetrad}) and (\ref{24.2}) we have $\omega_{\hat{1}\hat{2}} = -\,{m
\over 2a}\,\sqrt{\sigma\over k + \sigma}$]. In its turn, the direction of 
the magnetic field (sign of $B_{0}$) depends on the sign of the charge $e$. 

With the help of (\ref{34}) the equations (\ref{11}) are reduced to 
the following two equations which describe the state of matter:
\begin{eqnarray}
p(t) &=& - {1\over 8b\kappa } + {B^{2}_{0}\over 2a^{4}} + 
\left({\lambda_{1} - 4\lambda_{3}\over 6}\right){\tau^{2}_{0}\over a^{6}},
\label{35.1}\\
\epsilon (t) &=& {1\over 8b\kappa } + {3B^{2}_{0}\over 2a^{4}} + 
\left({\lambda_{1} - 4\lambda_{3}\over 6}\right){\tau^{2}_{0}\over a^{6}},
\label{35.2}
\end{eqnarray}
Finally, the evolution of the scale factor $a(t)$ is given by the equation
(\ref{30}). Substituting (\ref{tetrad}), (\ref{24.2}), and (\ref{25}) in it, 
one finds
\begin{equation}
{\ddot{a}\over a} + {\dot{a}^{2}\over a^{2}} - 
{m^{2}\left(3\sigma + 4k\right)\over 12ka^{2}} - \left({k+\sigma \over 36k}
\right)\left(4\lambda ^{2}_{3} - \lambda ^{2}_{1}\right)
{\tau^{2}_{0}\over a^{6}} - {k+ \sigma \over 12kb} =0.\label{36}
\end{equation}
As one can see, depending on the values of the coupling constants
$\lambda_{3}$ and $\lambda _{1}$, torsion can either accelerate or 
prevent the cosmological collapse. We will consider the latter possibility,
assuming, for concreteness, that $4\lambda ^{2}_{3}\gg\lambda ^{2}_{1}$.

Integrating (\ref{36}), one gets
\begin{equation}
\dot{a}^{2} = {m^{2}(3\sigma +4k)\over 12k} + {a^{2}(k+\sigma)\over 24kb} 
- {(k+\sigma )\left({4\lambda^2_3 - \lambda^2_1}\right){\tau}_0^2
\over 36ka^{4}} + {\alpha \over a^{2}}.\label{37}
\end{equation}
The integration constant $\alpha$ is determined by the normalization
of the scale factor $a = 1$ at the moment of a bounce ($t =0$), when
$\dot{a} =0$,
\begin{equation}
\alpha  = \left({k+ \sigma \over 36k}\right)\left(4\lambda^{2}_{3} - 
\lambda^{2}_{1}\right)\tau^{2}_{0} - {m^{2}\left(3\sigma +4k\right)
\over 12k} - {k+ \sigma \over 24kb} .\label{38}
\end{equation}

\subsection{Estimates of parameters of the cosmological model}

In order to integrate the equation (\ref{37}), one needs to know the 
constant parameters which enter that equation, because their numeric 
essentially affect the form of solution. All constants are uniquely
determined by the data which describe modern state of the universe: the 
equation of state $p_{t}\approx 0$ (dust), energy density $\epsilon_{t}$, 
Hubble constant $H_{t}(= \dot{a}/a)$ and the deceleration parameter 
$q_{t}\left(= -\,\ddot{a}a/\dot{a}^{2}\right)$ are estimated by
\begin{eqnarray*}
0.15 < &{\kappa \epsilon _{t}\over H^{2}_{t}}& <  12,\\
4\cdot 10^{-11}\,{\rm yr}^{-1} < &H_{t}& < 10^{-10}\,{\rm yr}^{-1}, \\
0.01 < &q_{t}& <  1.
\end{eqnarray*}
(Hereafter the index $t$ denotes modern values of the physical quantities).

Previously, \cite{birch1}, the angular velocity was estimated as relatively 
small $\omega _{t}/H_{t}\approx 10^{-3}$. New analyses of observational data 
suggests higher estimates for the global vorticity, more close to $H_{t}$. 
For the qualitative analysis of the model let us choose
\begin{equation}
q_{t}= 0.01,\qquad {\kappa \epsilon _{t}\over H^{2}_{t}} = 0.18,\qquad 
{\omega _{t}\over H_{t}} = 0.1.\label{39}
\end{equation}
Besides the estimates (\ref{39}), one needs also the values of the 
spin-torsion coupling constants. There is a large ambiguity which affects
the scale of the space at the moment of a bounce. If one assumes, cf. 
Trautman \cite{traut}, that the cosmological collapse stops on the scale
of order of 1 cm, then we get from (\ref{39}) an estimate
\begin{equation}
{\kappa \over \lambda_1 - 4\lambda_3}\approx 2.3\cdot 10^{27}.\label{40}
\end{equation}
Comparison of the equations (\ref{36})-(\ref{38}) with (\ref{39})-(\ref{40})
yields the modern value of the spin density
\begin{equation}
3\cdot 10^{8}\,{\rm g\,cm}^{-1}\,{\rm s}^{-1} \, < \tau_{t} <
7.6\cdot 10^{8}\,{\rm g\,cm}^{-1}\,{\rm s}^{-1} ,\label{41}
\end{equation}
the geometric parameters of the model
\begin{equation}
{k\over\sigma}\approx  71,\label{42}
\end{equation}
the coupling constant
\begin{equation}
b^{-1}= 0.36\,H^{2}_{t},\label{43}
\end{equation}
and the magnetic field
\begin{equation}
10^{-6} {\rm G} < B_{t} <   2.6\cdot 10^{-6} {\rm G}.\label{44}
\end{equation}
The final integration of the equation (\ref{37}) determines the law of
evolution of the scale factor:
\begin{equation}
4\sqrt{\alpha_3}\,H_{t}\,t = ({\rm const}) + {\ln Z}(a^{2}) 
+ {2\mu \over \widehat{k}\sqrt{-B_1B_2}}\left[{\gamma _{1}\over \gamma _{1}
+\gamma _{2}}F(\varphi ;\widehat{m}) - C_{1}\Pi (\varphi ;n,\widehat{m})
\right] ,\label{45}
\end{equation}
where the integration constant (const) is such that $a =1$ for $t=0$, 
and we denote the function of the scale factor 
$$
Z(a^{2}) = {(a^{2}+\gamma _{1})^{2}+\sqrt{a^2(a^2 -1)(a^2 - u_1)(a^2 - u_2)}
\over (\gamma _{1}+\gamma _{2})(\gamma _{1}-\gamma _{2} + 2a^{2})} - 
{C_{1}+C_{2}\over 2} .
$$

In eq. (\ref{45}), the constant parameters are determined by:
$$
\alpha_{1}= {m^{2}(3\sigma +4k)\over 12kH^{2}_{t}},\qquad 
\alpha_{2}= \left({k+\sigma \over k}\right)\left({4\lambda^{2}_{3} - 
\lambda^{2}_{1}\over 36}\right){\tau^{2}_{0}\over H^{2}_{t}},\qquad 
\alpha_{3}= {k+\sigma \over 12bkH^2_t} ;
$$
$u_{1}>u_{2}$ are the roots of the quadratic equation
$$
u^{2} + \left(1 + {\alpha_{1}\over \alpha_{3}}\right)u + 
{\alpha_{2}\over \alpha_{3}} = 0 ,
$$
from which one constructs
$$
\lambda_\pm = - u_{1}u_{2}\left(\sqrt{1-1/u_1}\pm\sqrt{1 - 1/u_2}\right)^2,
$$
$$
B_{1}= {\lambda _{+}\left(1-\lambda _{-}\right)\over \lambda _{+}-
\lambda _{-}},\qquad C_{1}= {\lambda _{-}\left(\lambda _{+}-1\right)\over 
\lambda _{+}-\lambda _{-}},\qquad B_{2}= {1- \lambda _{-}\over \lambda _{+}
-\lambda _{-}},\qquad C_{2}= {\lambda _{+} -1\over \lambda _{+}-\lambda _{-}},
$$
$$
\gamma_{1}= \sqrt{u_{1}u_{2}\over 1-\lambda_{-}},\qquad 
\gamma_{2}= \sqrt{u_{1}u_{2}\over 1-\lambda_{+}},\qquad \mu^{2}
= - {B_{1}\over C_{1}},\qquad \widehat{k}^{2}= {\lambda_{+}\over \lambda_{-}} ,
$$
$$
\widehat{m}^{2}= \left(\lambda_{+} - \lambda_{-}\right)/\lambda_{+},\qquad 
n = B_{1}\widehat{m}^{2}= 1 - \lambda_{-}.
$$
Finally, $F\left(\varphi;\widehat{m}\right)$ and $\Pi(\varphi;n,
\widehat{m})$ are elliptic integrals of the first and the third kinds,
respectively, with their argument given by
$$
\left(1 - \widehat{k}^{-2}\right)\sin^2\varphi  = 1 - 
\mu ^{-2}\left({a^{2}-\gamma _{2}\over a^{2}-\gamma _{1}}\right)^2 .
$$
The complete picture of the cosmological evolution is obtained when one
substitutes (\ref{39})-(\ref{44}) into the above equations. In particular,
one can verify that the age of the universe [estimated as the time since
$t=0$ till the modern stage characterized by (\ref{39})] is equal 
$T \approx  H^{-1}_{t}$, in a good agreement with observational data.

The analysis of the evolution of the scale factor $a(t)$, see (\ref{45}) and 
(\ref{37}), shows that one can naturally distinguish several qualitatively 
different stages in the history of the universe. First stage is the shortest 
one and it describes a bounce in the vicinity of $t=0$. There is no initial 
singularity due to the dominating spin contribution in (\ref{37}). At this
stage, the fluid source is characterized by the approximate equation of 
state of stiff matter 
$$
p = \epsilon\approx\left({\lambda_{1}-4\lambda_{3}\over 6}\right)
{\tau^2_0\over a^6}
$$
The applicability of the classical (non-quantum) gravitational theory is
guaranteed by the condition on the curvature invariant $\widetilde{R}
{}_{\alpha\beta\mu\nu}\widetilde{R}^{\alpha\beta\mu\nu}\ll l^{-4}_{\rm pl}$,
which is satisfied at the moment of the bounce. The duration of that stage
is quite small ($\ll 1\, s$), since the spin term quickly decreases in
(\ref{37}) with the growth of the scale factor.

Next comes the second stage when the scale factor increases like $\sqrt{t}$, 
while the equation of state is of the radiation type, $p\approx\varepsilon /3$.
This ``hot universe" expansion lasts until the size of the metagalaxy 
approaches $\approx 10^{27}$ cm. After this the ``modern'' stage starts with 
the effectively dust equation of state $p_{t}\approx 0,\,\epsilon_{t}\approx
(2b\kappa )^{-1}$. Scale factor still increases, but the deceleration of 
expansion takes place. The final stage depends on the value of the 
cosmological term, and either the future evolution enters the eternal
de Sitter type expansion, or expansion ends and a contraction phase starts. 

Our purpose in this section was to demonstrate the exact solution which
avoids the principal difficulties of old cosmological models with rotation
and expansion. Certainly, one cannot claim that the results obtained 
describe the real universe, and consequently we have limited ourselves to
a test type computations for the physical and geometrical parameters,
without trying to find the best estimates. The main difficulty of this 
particular dynamical realization, in our opinion, is presented by the 
magnitude of magnetic field (\ref{44}) which at the ``modern'' stage should 
be close to the upper limits established for the global magnetic field from 
astrophysical observations, see e.g. \cite{magnet}. 

It seems worthwhile to note that (\ref{41}) gives a rather big value for
the spin density ($\approx 10^{35}\hbar/{\rm cm}^{3}$). This is in a good
agreement with the macroscopic interpretation of the cosmological spin
fluid, the elements of which are spinning galaxies. Indeed, $\tau_{t}/
\rho_t\approx 10^{107}\hbar$ is close to the observed magnitude of the
angular momentum of a typical galaxy. Moreover, for the total spin of the
metagalaxy occupying the volume $V \approx  10^{84}{\rm cm}^{3}$, one finds 
$\tau_{\rm tot} = \tau_{t}V \approx 10^{120}\hbar$. This once again draws
attention to the Large Number hypothesis of Dirac-Eddington and related
questions, see \cite{brosche1,brosche2,brosche3,brosche4,largeN,z1}
and \cite{murad1,murad2,murad3,murad4}.

\section{Null geodesics in the G\"odel type model}

Practically all the information about the structure of the universe and about
the properties of astrophysical objects is obtained by an observer in the
form of different kinds of electromagnetic radiation. Thus, in order to be
able to make theoretical predictions and compare them with observations,
it is necessary to know the structure of null geodesics in the cosmological
model with rotation. All the models from the class (\ref{met0}) have
three Killing vectors and one conformal Killing vector field. Hence the
null geodesics equations (\ref{light}) 
have four first integrals,
\begin{equation}
q_0 =\xi^{\mu}_{\rm conf}k_{\mu}, \qquad 
q_a =\,-\,\xi^{\mu}_{(a)}k_{\mu}, \qquad a=1,2,3.\label{1int}
\end{equation}
Solving (\ref{1int}) with respect to $k^{\mu}$, one obtains a system of
ordinary first order nonlinear equations which can be straightforwardly 
integrated. Complete solution of the null geodesics equations in the 
G\"odel type model is given in \cite{prep,banach}, and here we present 
only short description of null geodesics in (\ref{met1}). 

To begin with, let us define convenient parametrization of null geodesics.
Without loosing generality (using the spatial homogeneity) we assume that an
observer is located at the space-time point $P=(t=t_0 , x=0, y=0, z=0)$.
Now, arbitrary geodesics which passes through $P$ is naturally determined by
its initial direction in the local Lorentz frame of observer at this point.
In the tetrad (\ref{tetrad}) we may put
\begin{equation}
k^{a}_P =(h^{a}_{\mu}\,k^{\mu})_P = (\,1,\,\sin\theta\cos\phi,\,
\sin\theta\sin\phi,\,\cos\theta ),\label{kini}
\end{equation}
where $\theta,\phi$ are standard spherical angles parametrizing the celestial 
sphere of an observer. Then from (\ref{conf}), (\ref{kil}), (\ref{kini}) and
(\ref{1int}) one finds the values of the integration constants
\begin{eqnarray}
q_0 &=& a_0,\nonumber\\
q_1 &=& a_0 (\sqrt{\sigma} + 
\sqrt{k+\sigma}\sin\theta\sin\phi),\label{const}\\
q_2 &=& a_0 \cos\theta,\nonumber\\
q_3 &=& {a_0\over m}\sin\theta\cos\phi.
\end{eqnarray}
Generic null geodesics, initial directions of which satisfy 
$q_1/a_0=(\sin\theta\sin\phi +\sqrt{\sigma\over{k+\sigma}})\neq 0$, 
are then described by 
\begin{eqnarray}
e^{-mx}&=&{{\sqrt{\sigma} + \sqrt{k+\sigma}\sin\theta\sin\Phi}\over
{\sqrt{\sigma} + \sqrt{k+\sigma}\sin\theta\sin\phi}},\label{x}\\
y&=&{{\sin\theta (\cos\Phi - \cos\phi )}\over 
{m(\sqrt{\sigma} + \sqrt{k+\sigma}\sin\theta\sin\phi)}},\label{y}\\
z&=&\left({{k+\sigma}\over k}\right)\cos\theta\left[
\int\limits^{t}_{t_0}{dt'\over a(t')} + \sqrt{\sigma\over{k+\sigma}}
\left({{\Phi - \phi}\over m}\right)\right],\label{z}
\end{eqnarray}
where the function $\Phi(t)$ satisfies the differential equation
\begin{equation}
{d\Phi\over dt}= -\,{m\over a}\,\left({
{\sqrt{\sigma\over {k+\sigma}} + \sin\theta\sin\Phi} \over
{1 + \sqrt{\sigma\over {k+\sigma}}\sin\theta\sin\Phi}}\right)\label{P}
\end{equation}
with initial condition $\Phi(t_0 )=\phi$. 

For a detailed discussion of rays which lie on the initial cone 
$(\sin\theta\sin\phi +\sqrt{\sigma\over{k+\sigma}})= 0$ and different 
subcases of (\ref{x})-(\ref{P}) see \cite{banach}. 

\section{Observations in rotating cosmologies}

The qualitative picture of specific rotational effects which could be 
observed in the G\"odel type model (\ref{met1}) is in fact independent of
the dynamical behaviour of the scale factor $a(t)$. To some extent the same
is true also for the quantitative estimates, especially if one uses the 
Kristian-Sachs formalism \cite{ks,mac3} in which all the physical and 
geometrical observable quantities are expressed in terms of power series in 
the affine parameter $s$ or the red shift $Z$. In this case the description
of observable effects on not too large (although cosmological) scales 
involves only the modern values (i.e. calculated at the moment of observation
$t=t_0$) of the scale factor $a_{0}=a(t_{0})$, Hubble parameter $H_{0}=
(\dot{a}/a)_{P}$, rotation value $\omega_{0}=\omega(t_{0})$, deceleration 
parameter $q_{0}=-(\ddot{a}a^2/\dot{a}^2)_P$, etc.  

In this section we will outline possible observational manifestations of 
the cosmic rotation in the G\"odel type universe. Estimates for the value
of vorticity and for the direction of rotation axis can be found from the
recent astrophysical data, see Sects.~\ref{period} and \ref{polarization}. 

\subsection{Kristian-Sachs expansions in observational cosmology}

Kristian and Sachs in the fundamental paper \cite{ks} have developed universal
technique for the analysis of the observational effects in an arbitrary 
cosmological model. Essentially, their method is based on the theory of null
geodesic (ray) bundles in Riemannian spacetime. The main advantage of the
Kristian-Sachs approach is in its generality which makes it possible to
consider cosmological tests even without the knowledge of explicit solution
of the gravitational field equations. This is particularly convenient when one 
is interested in revealing additional (as compared to the standard Hubble
expansion) geometrical and physical properties of the universe. In this
section we will apply the method of Kristian and Sachs to the rotating
cosmological models.

Let $k^\mu$ be a wave vector field which satisfies the null geodesics 
equation (\ref{light}) in a Riemannian manifold. (The connection is 
torsion-free, and we will omit the tilde to simplify the notation).
The starting point is the equation
\begin{equation}
\ddot{\Delta}^\mu\,=\,-\,R^\mu{}_{\alpha\nu\beta}\,\Delta^\nu\,k^\alpha\,
k^\beta,\label{deveq}
\end{equation}
which determines the deviation vector $\Delta^\mu:=\delta x^\mu$ of the
rays in a bundle. Hereafter dots denote covariant derivatives along a ray:
$(\,\dot{}\,)=k^\mu\nabla_\mu$, e.g., $\dot{\Delta}^\mu = D\Delta^\mu/ds =
k^\nu\nabla_\nu\Delta^\mu$, etc.

The crucial idea of Kristian and Sachs is to find a solution of (\ref{deveq})
in the form of a series in the affine parameter $s$. Correspondingly, all the
physical and geometrical quantities are ultimately represented by series in
$s$. Let us give some basic results of that approach, without going into
calculational details.

We will denote the spacetime points of an observer and a source by $P$ and
$S$, respectively. Let us start with the red shift defined by (\ref{red}).
A simple Taylor expansion of the right-hand side yields:
\begin{equation}
Z = \,-\,\overline{s}\,{\buildrel (1)\over u}\, + \,
{\overline{s}{}^2\over 2}\,{\buildrel (2)\over u}\,
-\,{\overline{s}{}^3\over 6}\,{\buildrel (3)\over u} + \dots\,.\label{Zs}
\end{equation}
Here we denoted:
$$
\overline{s}:= s\,(k^\mu u_\mu)_P,\qquad  
K^{\mu}=\left({k^{\mu}\over k^{\nu}u_{\nu}}\right), \qquad
{\buildrel (n)\over u} := \left(K^{\mu_1}\dots K^{\mu_{n-1}}K^{\mu_n}
\nabla_{\mu_1}\dots\nabla_{\mu_{n-1}}u_{\mu_n}\right)_P.
$$
Note that the value $s$ describes a position of the observer on a null geodesic
connecting $S$ and $P$, so that $s=0$ corresponds to $S$. The minus signs in
(\ref{Zs}) arise when we expand $(u^\mu k_\mu)_S$ at $P$. 

Let us consider a bundle of rays emitted from a point source $S$. We are
interested in $\Delta^\mu(s)$ along a geodesic from $S$ to $P$. One has
$\Delta^\mu(0)$, and hence from (\ref{deveq}), $\ddot{\Delta}^\mu(0)=0$.
The vector $\dot{\Delta}^\mu(0)$ is chosen to be orthogonal to $k^\mu$ and
$u^\mu$ at $S$, i.e. we are studying the cross-section of the bundle which 
is orthogonal to the central ray in the rest frame of the source. Note that
such an initial condition provides orthogonality $k^\mu\Delta_\mu =0$ at any
value of $s$, in view of (\ref{deveq}). 

The {\it area distance} is one of the central notions in the observational
cosmology. If an observer $P$ is viewing the distant source $S$ with an
intrinsic perpendicular area $dA_S$ which subtends the solid angle $d\Omega_P$
at $P$, the area distance $r$ between $P$ and $S$ is defined by
\begin{equation}
dA_S = r^2 d\Omega_P, \label{area}
\end{equation}
Another (reciprocal) area distance $r_S$ between $S$ and $P$ is defined by
$dA_P = r^2_S d\Omega_S$, where the solid angle $d\Omega_S$ subtends the
area $dA_P$ at $P$. The {\it reciprocity theorem} \cite{ellis1,ellis2,mac3}
tells that these distances are related by
\begin{equation}
r^2_S = r^2\,(1 + Z)^2.\label{recip}
\end{equation}
One can obtain $r_S$ with the help of $\Delta^\mu(s)$ which describes the
behaviour of the bundle of rays. 

The deviation equation (\ref{deveq}) determines iteratively the third and
higher covariant derivatives of $\Delta^\mu$ at $S$, and thus one construct
the solution in the form of the Taylor series
$$
\Delta^\mu(s) = s\,\dot{\Delta}^\mu(0) + 
{s^3\over 6}\,{\buildrel ... \over \Delta}{}^\mu(0) + \dots\, . 
$$
In order to find $r_S$, one needs actually only the length of $\Delta^\mu$.
Straightforward calculation yields:
\begin{eqnarray}
\Delta^\mu(s)\Delta_\mu(s) &=& s^2\,\dot{\Delta}^\mu(0)\dot{\Delta}_\mu(0)
\left[1 - {s^2\over 6}\,b - {s^3\over 12}\,\dot{b}+O(s^4)\right]_S\nonumber\\ 
&& \,+\,s^2\,\dot{\Delta}^\mu(0)\dot{\Delta}^\nu(0)\left[ - \,{s^2\over 3}\,
c_{\mu\nu} - {s^3\over 6}\,\dot{c}_{\mu\nu} + O(s^4)\right]_S .\label{del2}
\end{eqnarray}
The spacetime curvature contributes via the expressions
\begin{equation}
b:=R_{\mu\nu}\,k^\mu\,k^\nu,\qquad 
c_{\alpha\beta}:=\,C_{\alpha\mu\beta\nu}\,k^\mu\,k^\nu,\label{bc}
\end{equation}
higher order corrections are contained in the terms $O(s^4)$. 

After some algebra, one finds from (\ref{del2}): 
\begin{equation}
r^2_S = s^2\,(k^\mu u_\mu)_S\left[1 - {s^2\over 6}\,b - 
{s^3\over 12}\,\dot{b}+ O(s^4)\right]_S.\label{rs}
\end{equation}
Combining (\ref{red}), (\ref{recip}), and (\ref{rs}), one can easily derive 
the series which represents the area distance $r$ in terms of $s$. 

However, since the value of the affine parameter $s$ for the source $S$ and 
the observer $P$ cannot be measured directly, the series in powers of $s$ are
not particularly useful. Instead, Kristian and Sachs proposed to invert the
relation (\ref{Zs}), and then substitute $s=s(Z)$ in all other quantities. 
Direct calculation yields the following final series representation:
\begin{equation}
1 + Z = 1\,-\,r\,{\buildrel (1)\over u}\, + \,{r^2\over 2}\,{\buildrel (2)
\over u}\,-\,{r^3\over 6}\left({\buildrel (3)\over u} + {1\over 2}\,B\right)
+ \dots\,,\label{zr}
\end{equation}
where, replacing (\ref{bc}), $B:=(R_{\mu\nu}\,K^\mu\,K^\nu)_P$. Inverting 
(\ref{zr}), one finds
\begin{equation}
r\,=\, {Z\over - {\buildrel (1)\over u}}\left[ 1 - {Z\over 2}\,
{{\buildrel (2)\over u}\over{\buildrel (1)\over u}{}^2} + {Z^2\over 2}\left( 
{{\buildrel (2)\over u}{}^2\over {\buildrel (1)\over u}{}^4} - {1\over 6}\,
{{\buildrel (3)\over u}\over{\buildrel (1)\over u}{}^3} - {1 \over 12}
{B\over {\buildrel (1)\over u}{}^2}\right) + O(Z^3)\right].\label{rz}
\end{equation}

\subsection{Classical cosmological tests}

Classical cosmological tests, such as apparent magnitude -- red shift ($m-Z$),
number counts -- red shift ($N-Z$), angular size -- red shift relations, and
some other, reveal specific dependence of astrophysical observables on the 
angular coordinates ($\theta, \phi$) in a rotating world. Thus a careful 
analysis of the angular variations of empirical data over the whole celestial 
sphere is necessary. 

The knowledge of null geodesics enables one to obtain the explicit form of 
the area distance $r$ between an observer at a point $P$ and any source $S$.
Each ray (a
null geodesic) $x^\mu(s;p^A)$ passing through $P$ is labelled by the two 
parameters $p^A :=(\theta,\phi),\ A=1,2$ which have a simple meaning of the
spherical angles fixing the direction of ray at $P$. Thus in general, $r$ 
is a function of the direction of observation, that is $r=r(\theta,\phi)$.
Besides this, $r$ of course depends on the value of the affine parameter
$s$, or equivalently on the moment of $t$ at which source radiates a ray
detected by an observer at $t_0$. The area of the cross-section of the
bundle of rays is
\begin{equation}
dA_S = \sqrt{\det G_{AB}}\,d\theta\,d\phi, \qquad G_{AB} = 
g_{\mu\nu}\,{\partial x^\mu\over\partial p^A}\,{\partial x^\nu
\over\partial p^B}.
\end{equation}
Since the solid angle in (\ref{area}) is defined, as usually, by 
$d\Omega_P = \sin\theta\,d\theta\,d\phi$, one easily obtains the general
expression for the area distance between an observer $P$ and a source $S$
which lies on the geodesic labeled by the observational angles $(\theta,\phi)$
at the value $s$ of the geodetic parameter,
\begin{equation}
r(s;\theta,\phi)={\sqrt{\det G_{AB}(s;\theta,\phi)}\over \sin\theta}.
\end{equation}
The knowledge of the null geodesics (\ref{x})-(\ref{P}) makes it possible 
to calculate area distances in the G\"odel type universe explicitly.
For example, using (\ref{x})-(\ref{P}), one can find for the area distance 
{\it along the axis of rotation} the following exact result
\begin{equation}
r^2(t; \theta=0) = {\sin^2\left(\int\limits_{t_0}^{t}dt'\,\omega(t')\right)
\over \omega^{2}(t)}.\label{dist0}
\end{equation}
The rotational effects are maximal in the directions close to the $z$ axis, 
and quite remarkably the observations in the direction of rotation
can be described by simple and clear formulas. 

As for an arbitrary direction, exact formulas become very complicated and 
it is much more convenient to replace them by the Kristian-Sachs expansions. 
One can use the expansions (\ref{zr}) and (\ref{rz}) in the calculations of 
observable effects in rotating cosmologies: All the angular dependent 
rotational contributions are contained in the coefficients of these expansions.

For the G\"odel type cosmology (\ref{met1}) the classical ($m-Z$) and 
($N-Z$) relations read as follows \cite{banach,2,jetp1}.
\begin{description}
\item[Apparent magnitude $m$ versus red shift $Z$:]

The apparent magnitude of a source $S$ with the energy flux $L_P$, measured
at $P$, is defined by $m = - {5\over 2}\log_{10}\,L_P$. Given the total 
output of the source $L$, $L_S = L/4\pi$, the luminosity distance is 
$D^2:= L_S/L_P = r^2(1 + Z)^4$, see \cite{ellis1,ellis2}. Using (\ref{rz}),
one finds for the G\"odel type universe:
\begin{eqnarray}
m &=& M - 5\log_{10}H_{0} + 5\log_{10}Z + {5\over 2}\,(\log_{10}e)(1-q_{0})Z
\nonumber\\
&& -\,5\log_{10}\left(1 + \sqrt{\sigma\over{k+\sigma}}\sin\theta\sin\phi\right)
\nonumber\\
&& - {5\over 2}\,(\log_{10}e)\,{\omega_0\over H_0}\,Z\,\sin\theta\cos\phi
\left[{\sqrt{\sigma\over{k+\sigma}} + \sin\theta\sin\phi \over\left(1 + \sqrt{
\sigma\over{k+\sigma}}\sin\theta\sin\phi\right)^2}\right] + O(Z^2).\label{mz}
\end{eqnarray}
Here $M=-{5\over 2}\log_{10}L_{S}$ is the  absolute 
magnitude of a source with an intrinsic luminosity $L_S$.

\item[Number of sources $N$ versus red shift $Z$:]

Let us assume that the sources of electromagnetic radiation are distributed
homogeneously with the average density $n(t)$. Neglecting the evolution of
sources, one can count the number of images seen by an observer as follows. 
Consider the bundle with a solid angle $d\Omega$ detected by an observer at 
$P$. When moving along the central ray, the distance between the two 
infinitesimally close cross-sections is equal $d\ell = (k^\mu u_\mu)ds$,
and the volume between these sections is $dA\,d\ell$, with $dA = r^2d\Omega$.
Hence, the number of sources contained between these two sections is given
by $dN = - n_S(k^\mu u_\mu)_S\,r^2\,d\Omega\,ds$. Integrating, and using 
(\ref{zr})-(\ref{rz}): 
\begin{eqnarray}
{dN\over d\Omega} &=& {n_{0}Z^3 \over {3H_{0}^{3}\left(1 + 
\sqrt{\sigma\over{k+\sigma}}\sin\theta\sin\phi\right)^3}}\Bigg[ 1 -
{3\over 2}(1 + q_{0})Z \nonumber\\
&& - 3{\omega_0 \over H_0}{\sin\theta\cos\phi
\left(\sqrt{\sigma\over{k+\sigma}} + \sin\theta\sin\phi\right)\over
\left(1 + \sqrt{\sigma\over{k+\sigma}}\sin\theta\sin\phi\right)^2} Z +
O(Z^2)\Bigg].\label{nz}
\end{eqnarray}
\end{description}

The ($N-Z$) relation (\ref{nz}) describes the number of sources per solid
angle $d\Omega$ observed up to the value $Z$ of red shift. One can estimate 
the global difference of the number of sources visible in two hemispheres of 
the sky, $N^{+}, N^{-}$, by integrating (\ref{nz}). The result is
\begin{equation}
{{N^{+}- N^{-}}\over{N^{+}+ N^{-}}}= {1\over 2}\sqrt{\sigma\over{k+\sigma}}
\left(3 - {\sigma\over{k+\sigma}}\right) + O(Z^2 ).\label{NN}
\end{equation}
It seems worthwhile to draw attention to the absence of a correction 
proportional to $Z$ in (\ref{NN}). It is difficult to make a comparison of 
our results with \cite{wesson,mavr,fen} as they study stationary rotating 
models in which there is no red shift. 

 For some time classical cosmological tests were carefully carried out for
 standard models, but later it was recognized that evolution of physical
 properties of sources often dominates over geometrical effects. However,
 specific angular irregularities predicted in rotating cosmologies, (\ref{mz}),
 (\ref{nz})-(\ref{NN}), may revive the importance of the classical tests. 

 \subsection{Periodic structure of the universe}\label{period}

Recent analysis of the large--scale distribution of galaxies \cite{bro1}
has revealed an apparently periodic structure of the number of sources as
a function of the red shift. Cosmic rotation may give a natural explanation
of this fact \cite{12}. The crucial point is the helicoidal behaviour of the 
null geodesics (\ref{x})-(\ref{z}) in the G\"odel type model in the directions 
close to the rotation axis. This yields a periodicity of the area distance as 
a function of red shift, and hence the visible distribution of sources turns 
out to be also approximately periodical in $Z$. 

This effect is most transparent for the direction of rays straight along
the axis of rotation. The area distance is then given by (\ref{dist0}). In
order to be able to make some quantitative estimates, let us assume the
polynomial law for the scale factor,
\begin{equation}
a(t)=a_{0}\left({t-t_\infty\over{t_0 - t_\infty}}\right)^\gamma,\label{scale}
\end{equation}
which is naturally arising in a number of cosmological scenarios ($0<\gamma
<1$). Then one can derive (analogously to (\ref{nz})) the distribution of 
number of sources per red shift per solid angle:
\begin{equation}
{dN\over{d\Omega dZ}}={n_{0}\over{\omega_{0}^{2}H_{0}(1+Z)^{1/\gamma}}}
\sin^2 \left({\gamma\omega_{0}\over{(1-\gamma)H_{0}}}\left[
(1+Z)^{(\gamma-1)/\gamma} - 1\right]\right).\label{pereff}
\end{equation}
Thus, the apparent distribution of visible sources is an oscillating function 
of red shift, with slowly decreasing amplitude. A similar generalized formula 
can be obtained for arbitrary directions, so that (\ref{pereff}) is modified 
by additional angular dependence of the magnitude of successive extrema of 
distribution function. 

The observational data \cite{bro1} give for the distance between maxima the 
value $128 h^{-1}$ Mps (where $H_{0}=100 h {\rm km\ sec}^{-1}{\rm Mps}^{-1}$). 
From this one can estimate the rotation velocity which is necessary to produce
such a periodicity effect,
\begin{equation}
\omega_{0}\approx 74 H_{0}.\label{omper}
\end{equation}
This result does not depend on $\gamma$.

\subsection{Polarization rotation effect}\label{polarization}

The cosmic rotation affects a polarization of radiation which propagates in 
the curved spacetime (\ref{met1}), and this produces a new observable effect 
which has been reported in the literature by Birch \cite{birch1,birch2} and
more recently by Nodland and Ralston \cite{nod1,nod2}. In the geometrical 
optics approximation, polarization is described by a space-like vector 
$f^{\mu}$ which is orthogonal to the wave vector, $f_{\mu}k^{\mu}=0$,
and is parallelly transported along the light ray, 
\begin{equation}
k^{\mu}\nabla_{\mu}f^{\nu}=0.\label{pol}
\end{equation}
The influence of the gravitational field of a compact rotating massive body 
on the polarization of light was investigated by many authors, see e.g., 
\cite{skr1,skr2,balazs,pleb,anile1,anile3,brans,fay,gnedin,kob,mash1}. The 
analysis of (\ref{pol}) for the rotating cosmologies shows that the global 
cosmic rotation also forces a polarization vector to change its orientation 
during the propagation along a null geodesics \cite{banach,mans,kuhne}. It is 
clear that this conclusion has physical meaning only when one can define a 
frame at any point of the ray with respect to which polarization rotates. 
Let us describe how this can be achieved.

As it is well known, gravitational field affects all the properties 
of an image  of a source, such as shape, size and orientation 
\cite{sachs,nf,ellis1,ellis2,seitz}. Like the rotation of the polarization 
vector, the deformation and rotation of an image depend both on the local 
coordinates and on the choice of an observer's frame of reference. However, 
if one considers a combination of the two problems, then a truly observable 
effect arises which is coordinate and frame independent. Putting it in another 
way, one should calculate the influence of the cosmic rotation on {\it the 
relative angle} $\eta$ between the polarization vector and the direction of 
a major axis of an image \cite{18}. This problem was discussed in the papers 
\cite{panov4,hasse2}, but in our opinion, their results are incomplete, in 
the sense that only null congruences converging at the point of observation 
were considered. 

The relative angle $\eta$ can be most conveniently defined within the 
framework of the Newman-Penrose spin coefficient formalism. Namely, it is 
sufficient to construct a null frame $\{l,n,m,\overline{m}\}$ is such a way 
that $l$ coincides with the wave vector $k$, and the rest of the vectors are 
covariantly constant along $l$. Then we can consider $m$ as a polarization 
vector, and thus a deformation of an image of a source, calculated with 
respect to the frame $\{l,n,m,\overline{m}\}$, gives the observable relative 
angle $\eta$. Let us describe explicitly the null frame:
\begin{eqnarray}
l &=& {a_0\over a}\Bigg[\left\{1 + \sqrt{\sigma\over {k+\sigma}}
\sin\theta\sin\Phi\right\}\partial_t \nonumber\\
&& +{1\over a}\sin\theta\cos\Phi\partial_x +
{e^{-mx}\over a\sqrt{k+\sigma}}\sin\theta\sin\Phi\partial_y +
{1\over a}\cos\theta\partial_z\Bigg],\label{ll}\\
n &=& {a\over a_0}\left({1\over{1+\cos\theta}}\right)
\left[\partial_t - {1\over a}\partial_z\right],\label{nn}\\
m &=&{e^{i\Psi}\over \sqrt{2}}\Bigg[ \left\{\sqrt{\sigma\over 
{k+\sigma}} + i\left({\sin\theta\over{1+\cos\theta}}\right)
e^{-i\Phi}\right\}\partial_t \nonumber \\
&& +{i\over a}\partial_x + {e^{-mx}\over a\sqrt{k+\sigma}}\partial_y -
{i\over a}\left({\sin\theta\over{1+\cos\theta}}\right)
e^{-i\Phi}\partial_z\Bigg],\label{m1}\\
\overline{m} &=& {e^{-i\Psi}\over \sqrt{2}}\Bigg[\left\{
\sqrt{\sigma\over {k+\sigma}} - i\left({\sin\theta\over{1+\cos\theta}}
\right)e^{i\Phi}\right\}\partial_t \nonumber\\
&& - {i\over a}\partial_x + {e^{-mx}\over a\sqrt{k+\sigma}}\partial_y +
{i\over a}\left({\sin\theta\over{1+\cos\theta}}\right)
e^{i\Phi}\partial_z\Bigg].\label{m2}
\end{eqnarray}
Here 
\begin{equation}
\Psi(t,z)=z{m\over 2}\sqrt{\sigma\over {k+\sigma}} + \Phi(t).\label{Psi}
\end{equation}
One should note that (\ref{ll})-(\ref{m2}) is a smooth field of frames which 
covers the whole spacetime manifold. Direct computation proves that (\ref{ll})
is the null geodesic congruence with an affine parametrization. We say
that this congruence is oriented along a direction given by the spherical
angles $(\theta,\phi)$ in the local Lorentz frame of an observer at $P$, 
because a null geodesics (\ref{x})-(\ref{P}) belongs to this congruence.

It is straightforward to find the spin coefficients (we are using the 
definitions of \cite{chandra}, and mark the spin coefficients with tildes 
in order to distinguish them from the other quantities in our paper):
\begin{eqnarray}
\widetilde{\varepsilon} &=& 0, \qquad\widetilde{\kappa} = 0,\qquad
\widetilde{\lambda} = 0, \qquad \widetilde{\nu} = 0,\label{ep-kap}\\
\widetilde{\rho} &=& - {a_0\over a^2}\Bigg[\dot{a}\left\{1 + 
\sqrt{\sigma\over {k+\sigma}}\sin\theta\sin\Phi \right\} 
+{m\over 2}\left({k\over{k+\sigma}}\right) {\sin\theta\cos\Phi\over
(1 + \sqrt{\sigma\over {k+\sigma}}\sin\theta\sin\Phi )}\Bigg] \nonumber\\
&& + i{a_0\over a^2}{m\over 2}\cos\theta\left(
{{\sqrt{\sigma\over {k+\sigma}} + \sin\theta\sin\Phi} \over
{1 + \sqrt{\sigma\over {k+\sigma}}\sin\theta\sin\Phi}}\right),\label{rho}\\
\widetilde{\sigma} &=&{a_0\over a^2}{m\over 2}\left({k\over{k+\sigma}}\right)
{e^{2i\Psi}\sin\theta \over {(1 + \sqrt{\sigma\over {k+\sigma}}
\sin\theta\sin\Phi )}} \Big(-\cos\Phi\Big[2\cos\theta - 1 \nonumber\\
&& - 2\cos^{2}\Phi (\cos\theta -1)\Big] + i\sin\Phi\left[\cos\theta - 
2\cos^{2}\Phi (\cos\theta -1)\right]\Big). \label{sigma}
\end{eqnarray}
We do not write the other spin coefficients, because their values are
irrelevant. Only one important step remains to be done: the spin coefficient 
$\widetilde{\pi}\neq 0$ in the frame (\ref{ll})-(\ref{m2}), and we need to
make an additional Lorentz transformation
\begin{equation}
l\longrightarrow l,\quad n\longrightarrow n + A^{*}m + A\overline{m} +
A^{*}A\ l, \quad m\longrightarrow m + Al, \quad \overline{m}\longrightarrow
\overline{m} + A^{*}l,\label{add}
\end{equation}
where the function $A$ satisfies equation $l^{\mu}\partial_{\mu}A +
\widetilde{\pi}^{*}=0$. This ensures that $\widetilde{\pi}=0$ in a new 
frame, while remarkably the transformation (\ref{add}) does not change any
of the spin coefficients (\ref{ep-kap})-(\ref{sigma}). Thus one obtains
finally the field of null frames $\{ l,n,m,\overline{m}\}$ with the required
properties: $l$ is the null geodesic congruence with affine parametrization,
while $n,m,\overline{m}$ are covariantly constant along $l$. The latter is
equivalent to $\widetilde{\kappa}=\widetilde{\varepsilon}=\widetilde{\pi}=0$.

As it is well known, deformation and rotation of an image along a null 
geodesics are described by the optical scalars \cite{sachs}:
\begin{equation}
\widetilde{\vartheta}=\,-\,{\rm Re}\widetilde{\rho},\qquad
\widetilde{\omega}={\rm Im}\widetilde{\rho},\qquad \widetilde{\sigma}.
\end{equation}
We now assume, for definiteness, that the polarization vector $f^\mu$ 
coincides with the vector $m^\mu$ of the above null frame. Let an image of
a source, as seen at the point corresponding to a value $s=s_1$ of the
affine parameter, be an ellipse with the major axis $p$ and the minor axis 
$q$. Then one can straightforwardly obtain \cite{18} for the angle of 
rotation of the major axis of the image at $s_2 =s_1 + \delta s$,
\begin{equation}
\delta\eta = -\,\widetilde{\omega}\delta s -  {{p^2 + q^2}\over{p^2 - q^2}}
\,{\rm Im}\widetilde{\sigma}\,\delta s.\label{eta}
\end{equation}
Integration along a ray gives the finite angle of rotation.  

It is worthwhile to notice that exactly along the cosmic rotation axis 
the observer at $P$ finds for the optical scalars
\begin{equation}
\widetilde{\vartheta}_{P}=H_{0},\qquad \widetilde{\omega}_{P}=\omega_{0},
\qquad \widetilde{\sigma}=0.
\end{equation}
Thus the effect of rotation of the polarization vector in this direction is
the most explicit.

As for an arbitrary direction of observation, we finally find from (\ref{eta})
for $k/\sigma\ll 1$ with the help of the Kristian--Sachs expansions 
(\ref{rs})-(\ref{rz}):
\begin{equation}
\eta = \omega_0\,r\,\cos\theta + O(Z^{2}).\label{final}
\end{equation}
This result is in a good agreement with the observational data reported 
\cite{birch1} on the dipole anisotropy of distribution of the difference 
between the position angles of elongation (the major axis) and polarization 
in a sample of 3CR radio sources. The estimate for the direction and the
magnitude of the cosmic rotation, obtained in \cite{2,jetp1,18} from 
Birch's data, reads
\begin{eqnarray}
l^{\circ} &=& 295^{\circ}\pm 25^{\circ},\qquad
b^{\circ}=24^{\circ}\pm 20^{\circ},\label{dir}\\
\omega_{0} &=& (1.8\pm 0.8)H_{0}.\label{om2}
\end{eqnarray}
Not entering into the discussion \cite{anod1,anod2,anod3,anod4} of the 
statistical significance of the results of Nodland and Ralston, we can make 
a comparison with the new data. In \cite{nod1}, the dipole effect of the 
rotation of the plane of polarization was reported in the form $\beta=
{1\over 2}\Lambda^{-1}_s\,r\,\cos\gamma$, where $\beta$
is the residual rotation angle between the polarization vector and the 
direction of the major axis of a source (remaining after the Faraday 
rotation is extracted), $r$ is the distance to the source, $\gamma$ is the
angle between the direction of the wave propagation and the constant 
vector $\vec{s}$. The analysis of the data for 160 radio sources yielded 
\cite{nod1} the best fit for the constant $\Lambda_s=(1.1\pm0.08)\,{2 
\over 3}\,{10^{15}\over H}\,$m yr$^{-1}$ (with the Hubble constant $H$), 
and the $\vec{s}$-direction RA\,(21h$\pm$2h), dec\,($0^\circ\pm20^\circ$).
In an attempt to explain their observations, Nodland and Ralston  
concluded that it is impossible to understand such an effect within the 
conventional physics. Instead, they considered a modified electrodynamics
with the Chern-Simons type term violating the Lorentz invariance \cite{car},
and related $\Lambda_s$ to the coupling constant of that term. However, 
interpreting the effect as arising from the cosmic rotation, we immediately 
find 
\begin{eqnarray}
l^{\circ} &=& 50^{\circ}\pm 20^{\circ},\qquad
b^{\circ}=-30^{\circ}\pm 25^{\circ},\label{dirNR}\\
\omega_0 &=& (6.5\pm 0.5)\,H_0.\label{omNR}
\end{eqnarray}
This is larger than the estimate (\ref{om2}) obtained from Birch's data 
\cite{birch1}. Also the direction of the cosmic vorticity is different from 
(\ref{dir}).
However, it is interesting to note that within the error 
limits, the two directions are orthogonal to each other. 

It is clear that further careful observations and statistical analyzes will 
be extremely important in establishing the true value of the cosmic rotation.
In particular, it has been recently claimed in \cite{anod4}, that the 
magnitude of the polarization rotation effect should be one or two orders
lower than reported in \cite{nod1}. The value of the cosmic vorticity then
should be reduced correspondingly. 

\section{Conclusions}

In our discussion of the properties of rotating cosmologies we have paid 
special attention to the G\"odel type model (\ref{met1}). However the main
conclusions are true also for the whole class of metrics (\ref{met0}). 
One can notice, that in some cases the numerical estimates obtained for the 
value of the vorticity do not agree with each other, e.g., see (\ref{omper}),
(\ref{om2}), and (\ref{omNR}). Such inconsistencies can reflect the fact 
that too few empirical data were analyzed until now, so that a further 
detailed discussion is required in order to obtain the final convincing 
estimates. Already now certain modifications of the above simple models 
are necessary: It is clear, in the light of the modern COBE results 
\cite{cobe1,cobe2,cobe3}, that the purely rotating models (\ref{met0}) 
should be replaced by the cosmologies with a nontrivial shear. A preliminary 
analysis of such generalizations shows that rotating models can be made 
compatible with the COBE data without destroying the rest of the rotational 
effects (in particular, without essential modification of the polarization 
rotation formulas). It may turn out, though, that some (or all) of the above 
mentioned effects are explained after all by some different physical (and 
geometrical) reasons, not related to the cosmic rotation. However, ``whether 
our universe is rotating or not, it is of fundamental interest to understand 
the interrelation between rotation and other aspects of cosmological models
as well as to understand the observational significance of an overall
rotation" \cite{ros}.

We believe that the cosmic rotation is an important physical effect which
should find its final place in cosmology. In this paper we outlined one
of the possible theoretical frameworks which can underlie our understanding
of this phenomenon.

\bigskip
{\bf Acknowledgments}. This work was partly supported by the Deutsche 
Forschungsgemeinschaft (Bonn) grant He 528/17-2. I thank Prof. K.-E.~Hellwig 
for the kind invitation to the Colloqium on Cosmic Rotation at the Technische 
Universit\"at Berlin. The warm hospitality of the organizers and highly
interesting discussions with the participants are gratefully acknowledged.


\end{document}